\documentstyle[preprint,aps,epsfig]{revtex}
\tighten
\begin{document}
\draft
\preprint{Preprint Numbers: \parbox[t]{45mm}{ANL-PHY-8788-TH-97\\
                                             nucl-th/9708029}}

\title{$\pi$- and $K$-meson Bethe-Salpeter Amplitudes}

\author{Pieter Maris and Craig D. Roberts} 
\address{Physics Division, Bldg. 203, Argonne National Laboratory,
Argonne IL 60439-4843}
\date{14/August/97}
\maketitle
\begin{abstract}
Independent of assumptions about the form of the quark-quark scattering
kernel, $K$, we derive the explicit relation between the flavour-nonsinglet
pseudoscalar meson Bethe-Salpeter amplitude, $\Gamma_H$, and the
dressed-quark propagator in the chiral limit.  In addition to a term
proportional to $\gamma_5$, $\Gamma_H$ necessarily contains qualitatively and
quantitatively important terms proportional to $\gamma_5\,\gamma\cdot P$ and
$\gamma_5\,\gamma\cdot k\,k\cdot P$, where $P$ is the total momentum of the
bound state.  The axial-vector vertex contains a bound state pole described
by $\Gamma_H$, whose residue is the leptonic decay constant for the bound
state.  The pseudoscalar vertex also contains such a bound state pole and, in
the chiral limit, the residue of this pole is related to the vacuum quark
condensate.  The axial-vector Ward-Takahashi identity relates these pole
residues; with the Gell-Mann--Oakes--Renner relation a corollary of this
identity.  The dominant ultraviolet asymptotic behaviour of the scalar
functions in the meson Bethe-Salpeter amplitude is fully determined by the
behaviour of the chiral limit quark mass function, and is characteristic of
the QCD renormalisation group.  The rainbow-ladder Ansatz for $K$, with a
simple model for the dressed-quark-quark interaction, is used to illustrate
and elucidate these general results.  The model preserves the one-loop
renormalisation group structure of QCD.  The numerical studies also provide a
means of exploring procedures for solving the Bethe-Salpeter equation without
a three-dimensional reduction.
\end{abstract}

\section{Introduction}
$\pi$- and $K$-mesons are the lightest hadrons and hence they play a
significant role in the phenomenology of low-to-intermediate energy nuclear
physics as mediators of the long-range part of the hadron-hadron interaction.
They are easily produced in electron-nucleon and nucleon-nucleon collisions
and therefore provide an ideal means of exploring models of hadronic
structure and subnucleonic degrees of freedom in nuclei.  As mesons, the
simple quark-antiquark valence-quark content of the $\pi$ and $K$ makes them
the simplest light-quark systems one can study as strong interaction bound
states; and this is a necessary step in developing a detailed understanding
of their properties and interactions in terms of the elementary degrees of
freedom in QCD.

Mesonic bound states are described by the homogeneous Bethe-Salpeter equation
[BSE], which is one of the Dyson-Schwinger equations~\cite{dserev} [DSEs]
characterising QCD.  The homogeneous BSE is an eigenvalue problem whose
eigenvalue is $P^2$, the square of the bound state mass, and whose
eigenvector is the bound state amplitude [fully amputated,
quark-antiquark-meson vertex].  This bound state, or Bethe-Salpeter,
amplitude is a crucial element in the calculation of production and
scattering processes involving mesons, as illustrated in
Refs.~\cite{dserev,pctrev,pich97}.  The BSE is familiar in the study of
scattering and binding in two-nucleon systems; and it is often illustrated,
and its features explored, via the problem of two elementary scalars
interacting via the exchange of a different elementary scalar~\cite{allbse}.
There have been many applications to the strong interaction meson spectrum,
with recent, extensive studies in this general framework being those of
Refs.~\cite{jm93,tt94,vary94}, which also cite related research.

Bethe-Salpeter equation studies can be characterised by their treatment of
the quark-antiquark scattering kernel, $K(q,k;P)$; a concrete calculation
being specified by a truncation of the skeleton expansion for $K$.  This
kernel also appears implicitly in the DSE for the dressed-quark propagator
[the QCD ``gap equation''] via the dressed-quark-gluon vertex,
$\Gamma^a_\nu(q,p)$.  In studies of the spectrum and interactions of bound
states of light-quarks, where dynamical chiral symmetry breaking [DCSB] and
Goldstone's theorem are particularly important, it is crucial to ensure that
$K$ and $\Gamma^a_\nu$ are ``mutually consistent'', by which we mean that
they must be such as to guarantee the preservation of the axial-vector
Ward-Takahashi identity~\cite{mrt97}.  Otherwise, as discussed and
exemplified in Refs.~\cite{tt94,vary94,gross94}, a qualitatively correct
description of the light-quark meson spectrum is not possible; i.e.,
``fine-tuning'' is necessary to properly describe the {\it theoretical ideal}
of the chiral limit, and the observational fact that the pion is
so-much-lighter than the characteristic hadronic scale: $m_\rho/2 \simeq
m_N/3\doteq M_q$, the constituent quark mass, but $m_\pi/2 \simeq 0.2 M_q$.

The rainbow-ladder truncation of the quark-DSE and meson-BSE, without a
three-dimensional reduction, is a specification of $\Gamma^a_\nu$ and $K$
that ensures the preservation of the axial-vector Ward-Takahashi identity.
It is fully specified by an Ansatz for the dressed-quark-quark interaction
and allows a qualitatively and quantitatively good description of
flavour-nonsinglet pseudoscalar, vector and axial vector mesons without
fine-tuning, even in very simple models~\cite{sep97}.  As such it is a
phenomenologically efficacious tool in this sector.

The fact that it describes the flavour-singlet pseudoscalar and scalar mesons
poorly is not often mentioned.  However, this defect is not crucial now that
its source has been identified and understood~\cite{brs96,cdr}.  Employing a
straightforward and systematic procedure for extending the rainbow-ladder
truncation, a procedure that preserves the axial-vector Ward-Takahashi
identity at every order, allows one to analyse the attractive and repulsive
terms order-by-order beyond ladder-truncation in the BSE.  One finds, for
example, that in the flavour-nonsinglet pseudoscalar channel the repulsive
terms are approximately cancelled by attractive terms that arise at the same
order, which explains why the ladder truncation provides a good approximation
in this channel.  This is not the case in the scalar channel where higher
order terms do not cancel in a like manner but lead to a net repulsive
effect, whose magnitude cannot be estimated {\it a priori}$\,$ but which
entails that the ladder truncation provides a poor approximation.  In the
flavour-singlet pseudoscalar channel, timelike gluon exchange diagrams arise
when one improves upon the ladder-truncation and these provide a plausible
mechanism for splitting the flavour-singlet and flavour-nonsinglet mesons;
i.e., for generating a significant $\eta$-$\eta^\prime$ mass
splitting~\cite{ua1}.  These observations illustrate and emphasise that the
rainbow-ladder truncation can lead to qualitatively and quantitatively
reliable conclusions if used judiciously.

Our goal herein is to provide a concrete illustration of the general results
of Ref.~\cite{mrt97}; i.e., of the importance, feasibility and essential
consequences of preserving the axial-vector Ward-Takahashi identity in BSE
studies of quark-antiquark bound states, and the extension of these results
to $SU(N_f\geq 3)$.  We employ a renormalisable DSE-model of QCD that preserves
the one-loop renormalisation characteristics of the dressed-quark and -gluon
propagators and the quark-gluon vertex.  This allows an explicit
demonstration of the renormalisation group flow of the vacuum quark
condensate, for example, and renormalisation point independence of physical
observables in this framework.  We concentrate on the $\pi$- and $K$-mesons
since this subsystem has all the complexity necessary for a complete
discussion of the features we wish to elucidate.

In Sec.~II we describe the DSE for the renormalised dressed-quark propagator;
this propagator being a critical element in the construction of the kernel in
the BSE for meson bound states.  We discuss this BSE in Sec.~III, along with
the constraints entailed by preserving the axial-vector Ward-Takahashi
identity.  In Sec.~IV we report a model study of the quark DSE and meson BSE,
and a range of $\pi$- and $K$-meson observables, illustrating the
model-independent results derived in the preceding sections.  We summarise
and conclude in Sec.~V.

\section{Quark Dyson-Schwinger Equation}
In a Euclidean space formulation, with
$\{\gamma_\mu,\gamma_\nu\}=2\delta_{\mu\nu}$, $\gamma_\mu^\dagger =
\gamma_\mu$ and $a\cdot b=\sum_{i=1}^4 a_i b_i$, the DSE for the
renormalised dressed-quark propagator is
\begin{eqnarray}
\label{gendse}
S(p)^{-1} & = & Z_2 (i\gamma\cdot p + m_{\rm bm})
+\, Z_1\, \int^\Lambda_q \,
g^2 D_{\mu\nu}(p-q) \frac{\lambda^a}{2}\gamma_\mu S(q)
\Gamma^a_\nu(q,p) \,,
\end{eqnarray}
where $D_{\mu\nu}(k)$ is the renormalised dressed-gluon propagator,
$\Gamma^a_\nu(q;p)$ is the renormalised dressed-quark-gluon vertex, $m_{\rm
bm}$ is the $\Lambda$-dependent current-quark bare mass that appears in the
Lagrangian and $\int^\Lambda_q \doteq \int^\Lambda d^4 q/(2\pi)^4$ represents
mnemonically a {\em translationally-invariant} regularisation of the
integral, with $\Lambda$ the regularisation mass-scale.  The final stage of
any calculation is to remove the regularisation by taking the limit $\Lambda
\to \infty$.  The quark-gluon-vertex and quark wave function renormalisation
constants, $Z_1(\mu^2,\Lambda^2)$ and $Z_2(\mu^2,\Lambda^2)$ respectively,
depend on the renormalisation point and the regularisation mass-scale, as
does the mass renormalisation constant $Z_m(\mu^2,\Lambda^2) \doteq
Z_2(\mu^2,\Lambda^2)^{-1} Z_4(\mu^2,\Lambda^2)$.  In Eq.~(\ref{gendse}), $S$,
$\Gamma^a_\mu$ and $m_{\rm bm}$ depend on the quark flavour, although we have
not indicated this explicitly.  However, in our analysis we assume, and
employ, a flavour-independent renormalisation scheme and hence all the
renormalisation constants are flavour-independent.

\subsection{General remarks about renormalisation}
\subsubsection{Dressed-quark propagator}
\label{secdqp}
The solution of Eq.~(\ref{gendse}) has the general form
\begin{equation}
\label{sinvp}
S(p)^{-1} = i \gamma\cdot p A(p^2,\mu^2) + B(p^2,\mu^2)
        = \frac{1}{Z(p^2,\mu^2)}\left[ i\gamma\cdot p + M(p^2,\mu^2)\right]\,,
\end{equation}
renormalised such that at some large\footnote{Herein, by ``large'' we mean
$\mu^2$ very-much-greater than the renormalisation-group-invariant
current-quark mass for the $s$-quark so as to ensure that, in our model
calculations, the renormalisation constants are flavour-independent to better
than 1\%.  It is possible to employ a modified subtraction scheme in which
the renormalisation constants are exactly flavour-independent, however, it
does not quantitatively affect our results and hence is an unnecessary
complication~\cite{pt84}.} spacelike-$\mu^2$
\begin{equation}
\label{renormS}
\left.S(p)^{-1}\right|_{p^2=\mu^2} = i\gamma\cdot p + m(\mu)\,,
\end{equation}
where $m(\mu)$ is the renormalised quark mass at the scale $\mu$.  In the
presence of an explicit, chiral symmetry breaking, current-quark mass one has
$Z_4 m(\mu) = Z_2 m_{\rm bm}$, neglecting O$(1/\mu^2)$ corrections associated
with dynamical chiral symmetry breaking that are intrinsically
nonperturbative in origin.

Multiplicative renormalisability in QCD entails that
\begin{equation}
\frac{A(p^2,\mu^2)}{A(p^2,\bar\mu^2)}
= \frac{Z_2(\mu^2,\Lambda^2)}{Z_2(\bar\mu^2,\Lambda^2)}
= A(\bar\mu^2,\mu^2) 
= \frac{1}{A(\mu^2,\bar\mu^2)}\,.
\end{equation}
Such relations can be used as constraints on model studies of
Eq.~(\ref{gendse}).  Explicitly, at one-loop order in perturbation theory,
\begin{equation}
\label{z2mu}
Z_2(\mu^2,\Lambda^2) 
= \left[\frac{ \alpha(\Lambda^2) }{\alpha(\mu^2)}
        \right]^{-\frac{\gamma_F}{\beta_1}}\,,
\end{equation}
where $\gamma_F= \case{2}{3}\xi$ and $\beta_1= N_f/3 - 11/2$, with $\xi$ the
gauge parameter and $N_f$ the number of active quark flavours.  At this
order,
\begin{equation}
\label{alphaq2}
\alpha(Q^2) = \frac{\pi}
             {-\case{1}{2} \beta_1 
                \ln\left[\frac{Q^2}{\Lambda_{\rm QCD}^2}\right]}\,.
\end{equation}
Clearly, at one-loop in Landau gauge [$\xi = 0$], $A(p^2,\mu^2)\equiv 1$, and
a deviation from this result in a solution of Eq.~(\ref{gendse}) is a
higher-loop effect. Such effects are always present in the self-consistent
solution of Eq.~(\ref{gendse}). 

The ratio $M(p^2,\mu^2)=B(p^2,\mu^2)/A(p^2,\mu^2)$ is independent of the
renormalisation point in perturbation theory; i.e., with $\mu\neq\bar\mu$,
\begin{equation}
\label{Mrpi}
M(p^2,\mu^2) = M(p^2,\bar\mu^2)\doteq M(p^2)\,, \; \forall\, p^2\,.
\end{equation}
At one-loop order:
\begin{equation}
\label{masanom}
m(\mu) \doteq M(\mu^2) = \frac{\hat m}
{\left(\case{1}{2}\ln\left[\frac{\mu^2}{\Lambda_{\rm QCD}^2}
                \right]\right)^{\gamma_m}}\,,
\end{equation}
where $\hat m$ is a renormalisation-point-independent current-quark mass and
$\gamma_m = 12/(33-2 N_f)$ is the anomalous dimension at this order; and
\begin{equation}
\label{zmdef}
Z_m(\mu^2,\Lambda^2) 
= \left[\frac{ \alpha(\Lambda^2) }{\alpha(\mu^2)}\right]^{\gamma_m}\,.
\end{equation}
In QCD, $\gamma_m$ is independent of the gauge parameter to all orders in
perturbation theory and the chiral limit is defined by $\hat m = 0$.
Dynamical chiral symmetry breaking is manifest when, for $\hat m=0$, one
obtains $m(\mu) \sim {\rm O}(1/\mu^2) \neq 0$ in solving Eq.~(\ref{gendse}),
which is impossible at any finite order in perturbation theory.\footnote{The
arguments presented herein cannot be applied in a straightforward fashion to
models whose ultraviolet behaviour is that of quenched QED$_4$, such as
Ref.~\cite{fr96}, where the chiral limit can't be defined in this way.  The
difficulties encountered in such cases are illustrated in Ref.~\cite{hsw97}.}
This is discussed and illustrated in Sec.~\ref{qdsesln}.

\subsubsection{Dressed-gluon propagator}
In a general covariant gauge the renormalised dressed-gluon propagator in
Eq.~(\ref{gendse}) has the general form
\begin{equation}
D_{\mu\nu}(k) = \left( \delta_{\mu\nu} - \frac{k_\mu k_\nu}{k^2}\right)
                \frac{d(k^2,\mu^2)}{k^2} + \xi\,\frac{k_\mu k_\nu}{k^4}\,,
\end{equation}
where $d(k^2,\mu^2) = 1/[1+\Pi(k^2,\mu^2)]$, with $\Pi(k^2,\mu^2)$ the
renormalised gluon vacuum polarisation.  The fact that the longitudinal
[$\xi$-dependent] part of $D_{\mu\nu}(k)$ is not modified by interactions is
the result of a Slavnov-Taylor identity in QCD: $k_\mu D_{\mu\nu}(k) = \xi\,
k_\nu/k^2$.  We note that Landau gauge is a fixed point of the
renormalisation group; i.e., in Landau gauge the
renormalisation-group-invariant gauge parameter is zero to all orders in
perturbation theory; hence we employ this gauge in all numerical studies herein.

Multiplicative renormalisability entails that
\begin{equation}
\frac{d(k^2,\mu^2)}{d(k^2,\bar\mu^2)}
= \frac{Z_3(\bar\mu^2,\Lambda^2)}{Z_3(\mu^2,\Lambda^2)}
= d(\mu^2,\bar\mu^2)
= \frac{1}{d(\bar\mu^2,\mu^2)}\,.
\end{equation}
At one-loop order in perturbation theory
\begin{equation}
\label{z3mu}
Z_3(\mu^2,\Lambda^2) 
= \left[\frac{ \alpha(\Lambda^2) }{\alpha(\mu^2)}
        \right]^{-\frac{\gamma_1}{\beta_1}}\,,
\end{equation}
where $\gamma_1 = \case{1}{3}N_f - \case{1}{4}(13 - 3\xi)$.

\subsubsection{Dressed-quark-gluon vertex}
The renormalised dressed-quark-gluon vertex in Eq.~(\ref{gendse}) is of the
form
\begin{equation}
\Gamma^a_\nu(k,p) = \frac{\lambda^a}{2} \Gamma_\nu(k,p)\,.
\end{equation}
As a fully amputated vertex, it is free of kinematic singularities.  The
general Lorentz structure of $\Gamma_\nu(k,p)$ is straightforward but
lengthy, involving $12$ distinct scalar form factors, and here we do not
reproduce it fully:
\begin{equation}
\Gamma_\nu(k,p) = \gamma_\nu F_1(k,p,\mu) + \ldots\,;
\end{equation}
but remark that Ref.~\cite{pt84}, pp.~80-83, and Refs.~\cite{bc80,vertex}
provide an elucidation of its structure, evaluation and properties.

Renormalisability entails that only the form factor $F_1$, associated with
the $\gamma_\nu$ tensor, is ultraviolet-divergent.  By convention, and
defining:
$
f_1(k^2,\mu^2)\doteq F_1(k,-k,\mu)\,,
$
$\Gamma_\nu(k,p)$ is renormalised such that at some large spacelike-$\mu^2$
\begin{equation}
f_1(\mu^2,\mu^2)= 1\,.
\end{equation}

Since the renormalisation is multiplicative, one has
\begin{equation}
\frac{f_1(k^2,\mu^2)}{f_1(k^2,\bar\mu^2)}
= \frac{Z_1(\mu^2,\Lambda^2)}{Z_1(\bar\mu^2,\Lambda^2)}
= f_1(\bar\mu^2,\mu^2)
= \frac{1}{f_1(\mu^2,\bar\mu^2)}\,.
\end{equation}
At one-loop in perturbation theory the vertex renormalisation constant is
\begin{equation}
\label{z1mu}
Z_1(\mu^2,\Lambda^2) = \left[
                \frac{\alpha(\Lambda^2)}{\alpha(\mu^2)}
                        \right]^{-\frac{ \gamma_\Gamma}{\beta_1}}\,,
\end{equation}
where $\gamma_\Gamma = \case{1}{2}[ \case{3}{4} (3+\xi) + \case{4}{3}\xi]$.

\subsection{Model for the quark DSE}
\label{modeldse}
In order to exemplify the results of Ref.~\cite{mrt97}, which we reiterate
and generalise in Sec.~\ref{chiralsym}, we must know the form of
$D_{\mu\nu}(k)$ and $\Gamma_\nu(k,p)$, not only in the ultraviolet where
perturbation theory is applicable, but also in the infrared, where
perturbation theory fails and lattice simulations are affected by
finite-volume artefacts.  $D_{\mu\nu}(k)$ and $\Gamma_\nu(k,p)$ satisfy DSEs.
However, studies of these equations in QCD are rudimentary and are presently
best used only to suggest qualitatively reliable Ans\"atze for these
Schwinger functions.  That is why all quantitative studies of the quark DSE
to date have employed model forms of $D_{\mu\nu}(k)$ and $\Gamma_\nu(k,p)$.

\subsubsection{Abelian approximation}
To introduce one commonly used pair of Ans\"atze, we use Eqs.~(\ref{z2mu}),
(\ref{z3mu}) and (\ref{z1mu}) and observe that 
\begin{equation}
\label{sumanom}
\frac{ 2 \gamma_F }{\beta_1} 
+ \frac{ \gamma_1} {\beta_1} -  \frac{2 \gamma_\Gamma}{\beta_1} = 1\,.
\end{equation}
Hence, on the kinematic domain for which $Q^2 \doteq (p-q)^2 \sim p^2\sim
q^2$ is large and spacelike, the renormalised dressed-ladder kernel in the
Bethe-Salpeter equation for the (fully-amputated) Bethe-Salpeter amplitude
behaves as follows:
\begin{eqnarray}
\lefteqn{g^2(\mu^2)\, D_{\mu\nu}(p-q) \,
\left[\rule{0mm}{0.7\baselineskip} \Gamma^a_\mu(p_+,q_+)S\,(q_+) \right] 
\times 
\left[ \rule{0mm}{0.7\baselineskip}S(q_-)\,\Gamma^a_\nu(q_-,p_-) \right] }\\
&& \nonumber  
= 4\pi\, \alpha(Q^2)\, D_{\mu\nu}^{\rm free}(p-q)\,
\left[\rule{0mm}{0.7\baselineskip}
        \frac{\lambda^a}{2}\gamma_\mu \,S^{\rm free}(q_+)\right]\times
\left[\rule{0mm}{0.7\baselineskip}S^{\rm free}(q_-)\,
        \frac{\lambda^a}{2}\gamma_\nu\right]\,,
\end{eqnarray}
where $P$ is the total quark-antiquark momentum, $p_+ \doteq p + \eta_P P$
and $p_- \doteq p - (1-\eta_P) P$ [see Eq.~(\ref{genbse})].  This
observation, and the intimate relation between the kernel of the pseudoscalar
BSE and the integrand in Eq.~(\ref{gendse})~\cite{brs96}, provides a means of
understanding the origin of an often used Ansatz for $D_{\mu\nu}(k)$; i.e.,
in Landau gauge, making the replacement
\begin{equation}
\label{abapprox}
g^2 D_{\mu\nu}(k) \to 4\pi\,\alpha(k^2) \,D_{\mu\nu}^{\rm free}(k)
\end{equation}
in Eq.~(\ref{gendse}), and using the ``rainbow approximation'':
\begin{equation}
\label{rainbow}
\Gamma_\nu(q,p)=\gamma_\nu \, .
\end{equation}

The Ansatz expressed in Eq.~(\ref{abapprox}) is often described as the
``Abelian approximation'' because the left- and right-hand-sides are {\it
equal} in QED.  In QCD, equality between the two sides of
Eq.~(\ref{abapprox}) cannot be obtained easily by a selective resummation of
diagrams.  As reviewed in Ref.~\cite{dserev}, Eqs.~(5.1) to (5.8), it can
only be achieved by enforcing equality between the renormalisation constants
for the ghost-gluon vertex and ghost wave function: $\tilde Z_1=\tilde Z_3$.

A mutually consistent constraint, which follows from $\tilde Z_1=\tilde Z_3$
at a formal level, is to enforce the Abelian Ward identity $Z_1 = Z_2$.  At
one-loop this corresponds to neglecting the contribution of the 3-gluon
vertex to $\Gamma_\nu$, in which case $\gamma_\Gamma \to \case{2}{3}\xi =
\gamma_F$.  This additional constraint provides the basis for extensions of
Eq.~(\ref{rainbow}); i.e., using Ans\"atze for $\Gamma_\nu$ that are
consistent with the vector Ward-Takahashi identity in QED~\cite{qedwti}, such
as Refs.~\cite{beyondbow,brw92,hawes94}.

The combination of Abelian and rainbow approximations [with $Z_1=1=Z_2$]
yields a mass function, $M(p^2)$, with the ``correct'' one-loop anomalous
dimension; i.e., $\gamma_m$ in Eq.~(\ref{masanom}) in the case of explicit
chiral symmetry breaking or $(1-\gamma_m)$ in its absence~\cite{aj88}.
However, other often used Ans\"atze for $\Gamma_\nu$~\cite{bc80,cp90} yield
different and incorrect anomalous dimensions for $M(p^2)$~\cite{fred}.  This
illustrates and emphasises that the anomalous dimension of the solution of
Eq.~(\ref{gendse}) is sensitive to the details of the asymptotic behaviour of
the Ans\"atze for the elements in the integrand.  One role of the
multiplicative renormalisation constant $Z_1$ is to compensate for this.

\subsubsection{Model for $g^2\,D_{\mu\nu}(p-q)\,\Gamma_\nu(q,p)$}
\label{secwmt}
Herein we employ a model for the kernel of Eq.~(\ref{gendse}) based on the
Abelian approximation:
\begin{equation}
\label{ouransatz}
Z_1\, \int^\Lambda_q \,
g^2 D_{\mu\nu}(p-q) \frac{\lambda^a}{2}\gamma_\mu S(q)
\Gamma^a_\nu(q,p)
\to
\int^\Lambda_q \,
{\cal G}((p-q)^2)\, D_{\mu\nu}^{\rm free}(p-q)
 \frac{\lambda^a}{2}\gamma_\mu S(q)
\frac{\lambda^a}{2}\gamma_\nu \,,
\end{equation}
with the specification of the model complete once a form is chosen for the
``effective coupling'' ${\cal G}(k^2)$.

One consideration underlying this Ansatz is that we wish to study subtractive
renormalisation in a DSE-model of QCD and it is not possible to determine
$Z_1$ without analysing the DSE for the dressed-quark-gluon vertex; a problem
we postpone.  Instead we explored various Ans\"atze for $\Gamma_\nu$ and
found that, with ${\cal G}(k^2)= 4\pi\alpha(k^2)$ for large-$k^2$, there was
always at least one Ansatz for $Z_1$ that led to the correct anomalous
dimension for $M(p^2)$.  This interplay between the the renormalisation
constant and the integral is manifest in QCD and Eq.~(\ref{ouransatz}) is a
simple means of implementing it.

In choosing a form for ${\cal G}(k^2)$ we noted that the behaviour of
$\alpha(k^2)$ in the ultraviolet; i.e., for $k^2> 1$-$2\,$GeV$^2$, is well
described by perturbation theory.  Constraints on the form of ${\cal G}(k^2)$
in the infrared come from the DSE satisfied by the dressed-gluon propagator,
$D_{\mu\nu}(k)$.  As summarised succinctly in Refs.~\cite{cdr,mrp},
qualitatively reliable studies of this equation indicate that the
dressed-quark-quark interaction is significantly enhanced in the infrared
such that on this domain it is well represented by an integrable
singularity~\cite{bp89}.  Combining these observations with
Eqs.~(\ref{sumanom})-(\ref{abapprox}), which illustrate the necessary
interplay between the anomalous dimensions of each term in the integrand of
Eq.~(\ref{gendse}), motivates the Ansatz
\begin{equation}
\label{gk2}
\frac{{\cal G}(k^2)}{k^2} =
8\pi^4 D \delta^4(k) + \frac{4\pi^2}{\omega^6} D k^2 {\rm e}^{-k^2/\omega^2}
+ 4\pi\,\frac{ \gamma_m \pi}
        {\case{1}{2}
        \ln\left[\tau + \left(1 + k^2/\Lambda_{\rm QCD}^2\right)^2\right]}
{\cal F}(k^2) \,,
\end{equation}
with ${\cal F}(k^2)= [1 - \exp(-k^2/[4 m_t^2])]/k^2$ and $\tau={\rm e}^2-1$.
(We use $N_f=4$ and $\Lambda_{\rm QCD}^{N_f=4}= 0.234\,{\rm GeV}$ in our
numerical studies.)  This is a simple modification of the form used in
Ref.~\cite{fr96}; one which preserves the one-loop renormalisation group
behaviour of QCD in the quark DSE.

The qualitative features of Eq.~(\ref{gk2}) are clear.  The first term is an
integrable infrared singularity~\cite{mn83} and the second is a finite-width
approximation to $\delta^4(k)$, normalised such that it has the same $\int
d^4k$ as the first term.  In this way we split the infrared singularity into
the sum of a zero-width and a finite-width piece.  The last term in
Eq.~(\ref{gk2}) is proportional to $\alpha(k^2)/k^2$ at large spacelike-$k^2$
and has no singularity on the real-$k^2$ axis.

There are ostensibly three parameters in Eq.~(\ref{gk2}): $D$, $\omega$ and
$m_t$ ($D= 2 \gamma_m m_t^2$ in Ref.~\cite{fr96}).  However, in our numerical
studies, using a renormalisation point $\mu=19\,$GeV, which is large enough
to be in the perturbative domain, we fixed $\omega=0.3\,$GeV$(=1/[.66\,{\rm
fm}])$ and $m_t=0.5\,$GeV$(=1/[.39\,{\rm fm}])$, and only varied $D$ and the
renormalised $u/d$- and $s$-current-quark masses in order to obtain a good
description of low-energy $\pi$- and $K$-meson properties.  As shown below, this
is achieved with
\begin{equation}
\label{params}
\begin{array}{ccc}
D= 0.781\,{\rm GeV}^2\,,\; &
m_{u/d}(\mu) = 3.74\,{\rm MeV}\,,\; &
m_s(\mu) = 82.5\,{\rm MeV}
\end{array}\,.
\end{equation}
(We do not consider isospin breaking effects herein.)  We chose the quoted
values of $\omega$ and $m_t$ primarily so as to ensure that ${\cal
G}(k^2)\approx 4\pi\alpha(k^2)$ for $k^2>2\,$GeV$^2$, as illustrated in
Fig.~\ref{gkplot}.  This is sufficient for our present illustrative study,
just as the form in Ref.~\cite{fr96} was sufficient therein.  However,
increasing sophistication and/or an exploration of a broader range of
observables is likely to require a more careful treatment of this or other
parametric forms.

Evolved according to Eq.~(\ref{masanom}), postponing until Sec.~\ref{qdsesln}
the discussion of whether this formula is appropriate, the ``best fit'' mass
values in Eq.~(\ref{params}) correspond to $m_{u/d}^{1 {\rm GeV}} =
6.4\,{\rm MeV}$ and $m_s^{1 {\rm GeV}} = 140\,{\rm MeV}$.  The ``best
fit'' values are sensitive to the behaviour of ${\cal G}(k^2)$ for $k^2 \sim
1$-$2\,$GeV$^2$, which can be illustrated by their dependence on $\omega$:
increasing $\omega \to 1.5\,\omega$, while maintaining a good fit to $\pi$-
and $K$-meson observables, requires a $\sim$10\% reduction in the value of
these masses.  With minor modifications of our parametrisation we can satisfy
our phenomenological constraints using current-quark masses that are a factor
of $\sim 1.5$-$2\,$smaller, as canvassed in Ref.~\cite{kim97}.  We would have
to apply tighter constraints in our phenomenological application to make a
statement about the current-quark masses of light-quarks that is more
accurate than this.  These considerations do not affect the ratio of our
fitted current-quark mass values, $m_s(\mu)/ m_{u/d}(\mu)=22.0$, which is
consistent with Refs.~\cite{jm93,sep97} and the discussion of
Ref.~\cite{kim97}.

\section{Pion and Kaon Bethe-Salpeter Equations}
The renormalised, homogeneous, pseudoscalar Bethe-Salpeter Equation (BSE) is
\begin{eqnarray}
\label{genbse}
\left[\Gamma_H(k;P)\right]_{tu} &= & 
\int^\Lambda_q  \,
[\chi_H(q;P)]_{sr} \,K^{rs}_{tu}(q,k;P)\,,
\end{eqnarray}
where: $H=\pi$ or $K$ specifies the flavour-matrix structure of the
amplitude; $\chi_H(q;P) \doteq {\cal S}(q_+) \Gamma_H(q;P) {\cal S}(q_-)$,
with ${\cal S}(q) = {\rm diag}(S_u(q),S_d(q),S_s(q))$; $q_+=q + \eta_P\, P$,
$q_-=q - (1-\eta_P)\, P$, with $P$ the total momentum of the bound state; and
$r$,\ldots,$u$ represent colour-, Dirac- and flavour-matrix indices.

In Eq.~(\ref{genbse}), $K^{rs}_{tu}(q,k;P)$ is the renormalised,
fully-amputated quark-antiquark scattering kernel, which also appears
implicitly in Eq.~(\ref{gendse}) because it is the kernel in the
inhomogeneous integral equation [DSE] satisfied by $\Gamma_\nu(q;p)$.
$K^{rs}_{tu}(q,k;P)$ is a $4$-point Schwinger function obtained as the sum of
a countable infinity of skeleton diagrams.  It is two-particle-irreducible,
with respect to the quark-antiquark pair of lines and does not contain
quark-antiquark to single gauge-boson annihilation diagrams, such as would
describe the leptonic decay of a pseudoscalar meson.\footnote{A connection
between the fully-amputated quark-antiquark scattering amplitude: $M = K + K
({\cal S}{\cal S}) K + \ldots\,$, and the Wilson loop is discussed in
Ref.~\cite{nora}.} The complexity of $K^{rs}_{tu}(q,k;P)$ is one reason why
quantitative studies of the quark DSE currently employ Ans\"atze for
$D_{\mu\nu}(k)$ and $\Gamma_\nu(k,p)$.  As illustrated by Ref.~\cite{mrt97},
however, the complexity of $K^{rs}_{tu}(q,k;P)$ does not prevent one from
analysing aspects of QCD in a model independent manner and proving general
results that provide useful constraints on model studies of QCD.

Equation~(\ref{genbse}) is an eigenvalue problem.  Solutions exist only for
particular, separated values of $P^2$; and the eigenvector associated with
each eigenvalue, the Bethe-Salpeter amplitude [BSA]: $\Gamma_H(k;P)$, is the
one-particle-irreducible, fully-amputated quark-meson vertex.  In the
flavour-octet channels the solutions with the lowest eigenvalues are the
$\pi$- and $K$-mesons.\footnote{We do not consider the $\eta$-meson because
of its mixing with the $\eta^\prime$, which cannot be described in ladder
approximation~\cite{sep97,brs96}; the truncation of the BSE employed in
Sec~\ref{secmodbse}.}  The solution of Eq.~(\ref{genbse}) has the general
form~\cite{LS69}
\begin{eqnarray}
\label{genpibsa}
\Gamma_H(k;P) & = &  T^H \gamma_5 \left[ i E_H(k;P) + 
\gamma\cdot P F_H(k;P) \rule{0mm}{5mm}\right. \\
\nonumber & & 
\left. \rule{0mm}{5mm}+ \gamma\cdot k \,k \cdot P\, G_H(k;P) 
+ \sigma_{\mu\nu}\,k_\mu P_\nu \,H_H(k;P) 
\right]\,,
\end{eqnarray}
where for bound states of constituents with equal current-quark masses the
scalar functions $E$, $F$, $G$ and $H$ are even under $k\cdot P \to - k\cdot
P$ and, for example, $T^{K^+} = \case{1}{2}\left(\lambda^4 + i
\lambda^5\right)$, with $\{\lambda^j, j=1\ldots 8\}$ the $SU(3)$-flavour
Gell-Mann matrices.  The requirement that the bound state contribution to the
fully-amputated quark-antiquark scattering amplitude: $M = K + K ({\cal
S}{\cal S}) K + \ldots\,$, have unit residue leads to the canonical
normalisation condition for the BSA:
\begin{eqnarray}
\label{pinorm}
\lefteqn{ 2 P_\mu = 
\int^\Lambda_q \left\{\rule{0mm}{5mm}
{\rm tr} \left[ 
\bar\Gamma_H(q;-P) \frac{\partial {\cal S}(q_+)}{\!\!\!\!\!\!\partial P_\mu} 
\Gamma_H(q;P) {\cal S}(q_-) \right]
\right. + 
} \\
& & \nonumber \left.  
\;\;\;\;\;\;\;\;\;\;\;\;\;\;\;\;\;\;\;
 {\rm tr} \left[ 
\bar\Gamma_H(q;-P) {\cal S}(q_+) \Gamma_H(q;P) 
        \frac{\partial {\cal S}(q_-)}{\!\!\!\!\!\!\partial P_\mu}\right]
\rule{0mm}{5mm}\right\} +  \\
& & \nonumber
\;\;\;\;\;\;\;\;\;\;
 \int^\Lambda_{q}\int^\Lambda_{k} \,[\bar\chi_H(q;-P)]_{sr} 
\frac{\partial K^{rs}_{tu}(q,k;P)}
{\!\!\!\!\!\!\!\!\!\!\!\!\partial P_\mu}\, 
[\chi_H(k;P)]_{ut}\,,
\end{eqnarray}
where $\bar \Gamma_H(k,-P)^{\rm t} = C^{-1} \Gamma_H(-k,-P) C$, with
$C=\gamma_2 \gamma_4$, the charge conjugation matrix, and $X^{\rm t}$
denoting the matrix transpose of $X$.

In Eq.~(\ref{genbse}), $E_H(k;P)\neq 0$ acts as a ``source'' in the equations
for $F_H(k;P)$, $G_H(k;P)$ and $H_H(k;P)$ so that, in general, these
subleading Dirac components of $\Gamma_H(k;P)$ are nonzero.

\subsection{Chiral symmetry}
\label{chiralsym}
In studies of flavour-octet pseudoscalar mesons a good understanding of
chiral symmetry, and its explicit and dynamical breaking, is crucial.  These
features are expressed in the renormalised axial-vector Ward-Takahashi
identity [AV-WTI]
\begin{equation}
\label{avwti}
-i P_\mu \Gamma_{5\mu}^H(k;P)  = {\cal S}^{-1}(k_+)\gamma_5\frac{T^H}{2}
+  \gamma_5\frac{T^H}{2} {\cal S}^{-1}(k_-) 
- M_{(\mu)}\,\Gamma_5^H(k;P) - \Gamma_5^H(k;P)\,M_{(\mu)} \,,
\end{equation}
where: $M_{(\mu)}= {\rm diag}(m_u(\mu),m_d(\mu),m_s(\mu))$; the renormalised
axial-vector vertex is given by
\begin{eqnarray}
\label{genave}
\left[\Gamma_{5\mu}^H(k;P)\right]_{tu} & = &
Z_2 \, \left[\gamma_5\gamma_\mu \frac{T^H}{2}\right]_{tu} \,+
\int^\Lambda_q \, [\chi_{5\mu}^H(q;P)]_{sr} \,K^{rs}_{tu}(q,k;P)\,,
\end{eqnarray}
with $\chi_{5\mu}^H(q;P) \doteq {\cal S}(q_+) \Gamma_{5\mu}^H(q;P) {\cal
S}(q_-)$; and the renormalised pseudoscalar vertex by
\begin{eqnarray}
\label{genpve}
\left[\Gamma_{5}^H(k;P)\right]_{tu} & = & 
Z_4\,\left[\gamma_5 \frac{T^H}{2}\right]_{tu} \,+ 
\int^\Lambda_q \,
\left[ \chi_5^H(q;P)\right]_{sr}
K^{rs}_{tu}(q,k;P)\,,
\end{eqnarray}
with $\chi_{5}^H(q;P) \doteq {\cal S}(q_+) \Gamma_{5}^H(q;P) {\cal S}(q_-)$.
Multiplicative renormalisability ensures that no new renormalisation
constants appear in Eqs.~(\ref{genave}) and (\ref{genpve})~\cite{mrt97,pw68}.

Any study whose goal is a unified understanding of the properties of
flavour-octet pseudoscalar mesons and other hadronic bound states must ensure
the preservation of the AV-WTI, which correlates the axial-vector vertex,
pseudoscalar vertex, bound state amplitudes and quark propagators.

\subsubsection{Chiral limit}
Equation~(\ref{avwti}) is valid for all values of the
renormalisation-group-invariant current-quark masses, in particular for the
chiral-limit when $M_{(\mu)}\Gamma_5^H(k;P) = {\rm diag(0,0,0)} =
\Gamma_5^H(k;P)\,M_{(\mu)}$.  In this case the AV-WTI is
\begin{equation}
\label{avwti0}
-i P_\mu \Gamma_{5\mu}^H(k;P)  = {\cal S}^{-1}(k_+)\gamma_5\frac{T^H}{2}
+  \gamma_5\frac{T^H}{2} {\cal S}^{-1}(k_-)\,.  
\end{equation}
As a straightforward generalisation of Ref.~\cite{mrt97}, it follows from
Eqs.~(\ref{sinvp}) and (\ref{avwti0}) that in the chiral limit the
axial-vector vertex has the form
\begin{eqnarray}
\label{truavv}
\Gamma_{5 \mu}^H(k;P) & = &
\frac{T^H}{2} \gamma_5 
\left[ \rule{0mm}{5mm}\gamma_\mu F_R(k;P) + \gamma\cdot k k_\mu G_R(k;P) 
- \sigma_{\mu\nu} \,k_\nu\, H_R(k;P) \right]\\
&+ & \nonumber
 \tilde\Gamma_{5\mu}^{H}(k;P) 
+\,f_H\,  \frac{P_\mu}{P^2 } \,\Gamma_H(k;P)\,,
\end{eqnarray}
where: $F_R$, $G_R$, $H_R$ and $\tilde\Gamma_{5\mu}^{H}$ are regular as
$P^2\to 0$; $P_\mu \tilde\Gamma_{5\mu}^{H}(k;P) \sim {\rm O }(P^2)$;
$\Gamma_H(k;P)$ is the pseudoscalar BSA in Eq.~(\ref{genpibsa}); and the
residue of the pseudoscalar pole in the axial-vector vertex is $f_H$, the
leptonic decay constant:
\begin{eqnarray}
\label{caint}
f_H P_\mu = 
Z_2\int^\Lambda_q\,\case{1}{2}
{\rm tr}\left[\left(T^H\right)^{\rm t} \gamma_5 \gamma_\mu 
{\cal S}(q_+) \Gamma_H(q;P) {\cal S}(q_-)\right]\,,
\end{eqnarray}
with the trace over colour, Dirac and flavour indices.  In addition the chiral
limit AV-WTI entails
\begin{eqnarray}
\label{bwti} 
f_H E_H(k;0)  &= &  B(k^2)\,, \\
\label{fwti}
 F_R(k;0) +  2 \, f_H F_H(k;0)                 & = & A(k^2)\,, \\
\label{rgwti}
G_R(k;0) +  2 \,f_H G_H(k;0)    & = & 2 A^\prime(k^2)\,,\\
\label{gwti} 
H_R(k;0) +  2 \,f_H H_H(k;0)    & = & 0\,,
\end{eqnarray}
where $A(k^2)$ and $B(k^2)$ are the solutions of Eq.~(\ref{gendse}) in the
chiral limit.

As remarked above, in perturbation theory, $B(k^2) \equiv 0$ in the chiral
limit.  The appearance of a $B(k^2)$-nonzero solution of Eq.~(\ref{gendse})
in the chiral limit signals DCSB: one has {\it dynamically generated} a
momentum-dependent quark mass term in the absence of a seed-mass.
Equations~(\ref{truavv}) and (\ref{bwti})-(\ref{gwti}) show that when chiral
symmetry is dynamically broken: 1) the homogeneous, flavour-nonsinglet,
pseudoscalar BSE has a massless, $P^2=0$, solution; 2) the BSA for the
massless bound state has a term proportional to $\gamma_5$ alone, with the
momentum-dependence of $E_H(k;0)$ completely determined by that of the scalar
part of the quark self energy, in addition to terms proportional to other
pseudoscalar Dirac structures, $F_H$, $G_H$ and $H_H$, that are nonzero in
general; and 3) the axial-vector vertex, $\Gamma_{5 \mu}^H(k;P)$, is dominated
by the pseudoscalar bound state pole for $P^2\simeq 0$.  The converse is also
true.

The relationship, in the chiral limit, between the normalisation of the
pseudoscalar BSA and $f_H$ has often been discussed; for example,
Refs~\cite{fr96,mmy88}.  Consider that if one chooses to normalise $\Gamma_H$
such that $E_H(0;0)= B(0)$, and defines the BSA so normalised as
$\Gamma_{H}^{N_H}(k;P)$, then the right-hand-side of (\ref{pinorm}),
evaluated with $\Gamma_H \to \Gamma_{H}^{N_H}(k;P)$, is equal to $2 P_\mu
N_H^2$, where $N_H$ is a dimensioned constant.  Using
Eqs.~(\ref{bwti})-(\ref{gwti}) it is clear that in the chiral limit
\begin{equation}
\label{npifpi}
N_H = f_H\,.
\end{equation}  
However, in model studies to date, this result is not obtained {\it unless}
one assumes $A(k^2)\equiv 1$.  It follows that any kernel which leads, via
(\ref{gendse}), to $A(k^2)\equiv 1$ must also yield, $F_H \equiv 0 \equiv
G_H\equiv H_H$, if it preserves the AV-WTI.  In realistic model studies,
where $A(k^2) \not \equiv 1$, the difference between the values of $N_H$
and $f_H$ is an artefact of neglecting $F_H$, $G_H$ and $H_H$ in
(\ref{genpibsa})~\cite{dserev}.

\subsubsection{Explicit chiral symmetry breaking}
Again as a straightforward generalisation of Ref.~\cite{mrt97}, in the
presence of explicit chiral symmetry breaking the AV-WTI, Eq.~(\ref{avwti}),
entails that both the axial-vector and the pseudoscalar vertices have a
pseudoscalar pole; i.e., 
\begin{eqnarray}
\label{truavvm}
\Gamma_{5 \mu}^H(k;P) & = &
\frac{T^H}{2} \gamma_5 
\left[ \gamma_\mu F_R^H(k;P) + \gamma\cdot k k_\mu G_R^H(k;P) 
- \sigma_{\mu\nu} \,k_\nu\, H_R^H(k;P) \right]\\
&+ & \nonumber
 \tilde\Gamma_{5\mu}^{H}(k;P) 
+\,f_H\,  \frac{P_\mu}{P^2 +m_H^2} \,\Gamma_H(k;P)\,,
\end{eqnarray}
and
\begin{eqnarray}
\label{trupvvm}
\Gamma_{5}^H(k;P) & = &
\frac{T^H}{2} \gamma_5 
\left[ i {\cal E}_R^H(k;P) + 
\gamma\cdot P\, {\cal F}_R^H(k;P) 
+ \gamma\cdot k\, k\cdot P\, {\cal G}_R^H(k;P) 
\right.\\
& & \nonumber
\left. + \sigma_{\mu\nu} \,k_\mu P_\nu\, {\cal H}_R^H(k;P) \right]
+\,r_H\,  \frac{1}{P^2 +m_H^2} \,\Gamma_H(k;P)\,,
\end{eqnarray}
with: ${\cal E}_R^H$, $F_R^H$, ${\cal F}_R^H$, $G_R^H$, ${\cal G}_R^H$,
$H_R^H$, ${\cal H}_R^H$ and $\tilde\Gamma_{5\mu}^{H}$ regular as $P^2\to
-m_H^2$; $P_\mu\tilde\Gamma_{5\mu}^{H}(k;P) \sim {\rm O}(P^2)$; and 
\begin{equation}
\label{gmora}
f_H\,m_H^2 = r_H \, {\cal M}_H\,,\;\;
{\cal M}_H \doteq {\rm tr}_{\rm flavour}
\left[M_{(\mu)}\,\left\{T^H,\left(T^H\right)^{\rm t}\right\}\right]\,,
\end{equation}
where $f_H$ is given by Eq.~(\ref{caint}), with massive quark propagators in
this case, and the residue of the pole in the pseudoscalar vertex is
\begin{equation}
\label{rH}
i r_H = Z_4\int^\Lambda_q\,\case{1}{2}
{\rm tr}\left[\left(T^H\right)^{\rm t} \gamma_5 
{\cal S}(q_+) \Gamma_H(q;P) {\cal S}(q_-)\right]\,.
\end{equation}
The factor $Z_4$ on the right-hand-side depends on the gauge parameter, the
regularisation mass-scale and the renormalisation point.  This dependence is
exactly that required to ensure that: 1) $r_H$ is finite in the limit
$\Lambda\to \infty$; 2) $r_H$ is gauge-parameter independent; and 3) the
renormalisation point dependence of $r_H$ is just such as to ensure that the
right-hand-side of Eq.~(\ref{gmora}) is renormalisation point {\it
independent}.  This is obvious at one-loop order, especially in Landau-gauge
where $Z_2\equiv 1$ and hence $Z_4 = Z_m$.

In the chiral limit, using Eqs.~(\ref{genpibsa}) and
(\ref{bwti})-(\ref{gwti}), Eq.~({\ref{rH}) yields
\begin{equation}
\label{cbqbq}
\begin{array}{lcr}
\displaystyle
r_H^0  =  -\,\frac{1}{f_H^0}\, \langle \bar q q \rangle_\mu^0 \,
, & & 
\displaystyle
\,-\,\langle \bar q q \rangle_\mu^0 \doteq  
Z_4(\mu^2,\Lambda^2)\, N_c \int^\Lambda_q\,{\rm tr}_{\rm Dirac}
        \left[ S_{\hat m =0}(q) \right]\,,
\end{array}
\end{equation}
where the superscript ``$0$'' denotes that the quantity is evaluated in the
chiral limit and $ \langle \bar q q \rangle_\mu^0 $, as defined here, is the
chiral limit {\it vacuum quark condensate}, which is renormalisation-point
dependent but independent of the gauge parameter and the regularisation
mass-scale.  Equation~(\ref{trupvvm}) is the statement that {\it the chiral
limit residue of the bound state pole in the flavour-nonsinglet pseudoscalar
vertex is} $(-\,\langle \bar q q \rangle_\mu^0)/f_H^0$.

From Eqs.~(\ref{gmora}) and (\ref{cbqbq}) one obtains immediately
\begin{eqnarray}
\label{gmorepi}
f_{\pi}^2 m_{\pi}^2 & = &-\,\left[m_u(\mu) + m_d(\mu)\right]
       \langle \bar q q \rangle_\mu^0 + {\rm O}\left(\hat m_q^2\right)\,\\
\label{gmoreKp}
f_{K^+}^2 m_{K^+}^2 & = &-\,\left[m_u(\mu) + m_s(\mu)\right]
       \langle \bar q q \rangle_\mu^0 + {\rm O}\left(\hat m_q^2\right)\,,
\end{eqnarray}
which exemplify what is commonly known as the Gell-Mann--Oakes--Renner
relation.  

We emphasise that the primary result, Eq.~(\ref{gmora}), of which
Eqs.~(\ref{gmorepi}) and (\ref{gmoreKp}) are corollaries, is valid {\it
independent} of the magnitude of $\hat m_q$.  We can rewrite it in the form
\begin{equation}
\label{gmorqbqM}
f_H^2\, m_H^2 = \,-\,\langle \bar q q \rangle_\mu^H\,{\cal M}_H
\end{equation}
where we have introduced the {\it notation}
\begin{equation}
\label{qbqM}
 -\,\langle \bar q q \rangle_\mu^H \doteq f_H\,r_H\,,
\end{equation}
in order to highlight the fact that, for nonzero current-quark masses,
Eq.~(\ref{gmora}) {\it does not}$\,$ involve a difference of vacuum
massive-quark condensates; a phenomenological assumption often
employed.\footnote{In QCD, the integral defining $r_H$ diverges
logarithmically, like the trace of the chiral limit quark propagator (vacuum
quark condensate), which is the reason why the right-hand-side of
Eq.~(\ref{gmora}) is independent of the renormalisation point.  This is
unlike the trace of the $\hat m \neq 0$ quark propagator, which diverges
quadratically.}

\subsection{Model BSE}
\label{secmodbse}
In order to exemplify the results of Ref.~\cite{mrt97}, which we have
reiterated and generalised in Sec.~\ref{chiralsym}, we must have an explicit
form for the kernel $K^{rs}_{tu}(q,k;P)$ in Eq.~(\ref{genbse}).  The form
must be such as to preserve the AV-WTI, Eq.~(\ref{avwti}), which requires a
truncation of the skeleton expansion for $K^{rs}_{tu}(q,k;P)$ that is
consistent with Eq.~(\ref{ouransatz}); our Ansatz for the kernel of
Eq.~(\ref{gendse}).  The ``ladder truncation'' fulfills this
requirement~\cite{brs96,cdr}:
\begin{equation}
K^{rs}_{tu}(q,k;P) = - \,{\cal G}((k-q)^2)\,D^{\rm free}_{\mu\nu}(k-q)\,
        \left(\gamma_\mu\frac{\lambda^a}{2}\right)_{tr}\,
        \left(\gamma_\nu\frac{\lambda^a}{2}\right)_{su}\,,
\end{equation}
in which case Eq.~(\ref{genbse}) becomes
\begin{equation}
\label{bsemod}
\Gamma_H(k;P) + \int^\Lambda_q
{\cal G}((k-q)^2)\, D_{\mu\nu}^{\rm free}(k-q)
 \frac{\lambda^a}{2}\gamma_\mu {\cal S}(q_+)\Gamma_H(q;P){\cal S}(q_-)
\frac{\lambda^a}{2}\gamma_\nu = 0\,,
\end{equation}
and the normalisation condition, Eq.~(\ref{pinorm}), simplifies because the
last term vanishes when $K^{rs}_{tu}(q,k;P)$ is independent of $P_\mu$.

\section{Numerical Results}
\subsection{Solution of the quark DSE}
\label{qdsesln}
Using Eqs.~(\ref{gendse}) and (\ref{ouransatz}) our model quark DSE is 
\begin{eqnarray}
\label{dsemod}
S(p,\mu)^{-1} & = & Z_2\, i\gamma\cdot p + Z_4\, m(\mu) 
+ \Sigma^\prime (p,\Lambda)\,,
\end{eqnarray}
with the regularised quark self energy
\begin{eqnarray}
\Sigma^\prime(p,\Lambda) & \doteq & \int^\Lambda_q \,
{\cal G}((p-q)^2)\, D_{\mu\nu}^{\rm free}(p-q)
 \frac{\lambda^a}{2}\gamma_\mu S(q)
\frac{\lambda^a}{2}\gamma_\nu \,,
\end{eqnarray}
where ${\cal G}(k^2)$ is given in Eq.~(\ref{gk2}).  Equation~(\ref{dsemod})
is a pair of coupled integral equations for the functions $A(p^2,\mu^2)$ and
$B(p^2,\mu^2)$ defined in Eq.~(\ref{sinvp}).

In the case of explicit chiral symmetry breaking, $\hat m \neq 0$, the
renormalisation boundary condition of Eq.~(\ref{renormS}) is straightforward
to implement.  Writing
\begin{equation}
\Sigma^\prime(p,\Lambda) \doteq i \gamma\cdot p \, 
                \left( A^\prime(p^2,\Lambda^2) - 1\right)
                        + B^\prime(p^2,\Lambda^2)\,,
\end{equation}
Eq.~(\ref{renormS}) entails
\begin{equation}
\label{z2def}
Z_2(\mu^2,\Lambda^2) = 2 - A^\prime(\mu^2,\Lambda^2)
\;\; {\rm and} \;\;
m(\mu) = Z_2(\mu^2,\Lambda^2)\,m_{\rm bm}(\Lambda^2) + 
        B^\prime(\mu^2,\Lambda^2)
\end{equation}
and hence
\begin{eqnarray}
\label{aren}
A(p^2,\mu^2) & = & 1 
        +  A^\prime(p^2,\Lambda^2) 
        - A^\prime(\mu^2,\Lambda^2)\,,\\
\label{bren}
B(p^2,\mu^2) & = & m(\mu) 
        +  B^\prime(p^2,\Lambda^2) 
        - B^\prime(\mu^2,\Lambda^2)\,.
\end{eqnarray}

From Sec.~\ref{secdqp}, having fixed the solutions at a single
renormalisation point, $\mu$, their form at another point, $\bar\mu$, is
given by
\begin{equation}
S^{-1}(p,\bar\mu) = i \gamma\cdot p\, A(p^2,\bar\mu^2) + B(p^2,\bar\mu^2)
        = \frac{Z_2(\bar\mu^2,\Lambda^2)}{Z_2(\mu^2,\Lambda^2)}
                S^{-1}(p,\mu)\,.
\end{equation}
[Recall that $M(p^2)$ is independent of the renormalisation point.]  This
feature is manifest in our solutions.  It means that, in evolving the
renormalisation point to $\bar\mu$, the ``1'' in Eq.~(\ref{aren}) is replaced
by $Z_2(\bar\mu^2,\Lambda^2)/Z_2(\bar\mu^2,\Lambda^2)$, and the ``$m(\mu)$''
in Eq.~(\ref{bren}) by $m(\bar\mu)$; i.e., the ``seeds'' in the integral
equation evolve according to the QCD renormalisation group.

As also remarked in Sec.~\ref{secdqp}, the chiral limit in QCD is
unambiguously defined by $\hat m = 0$.  In this case there is no perturbative
contribution to the scalar piece of the quark self energy, $B(p^2,\mu^2)$,
and, in fact, there is no scalar, mass-like divergence in the perturbative
calculation of the self energy.  It follows that $Z_2(\mu^2,\Lambda^2) m_{\rm
bm}(\Lambda^2)=0\,,\forall \Lambda$ and, from Eqs.~(\ref{z2def}) and
(\ref{bren}), that there is no subtraction in the equation for
$B(p^2,\mu^2)$; i.e., Eq.~(\ref{bren}) becomes
\begin{equation}
B(p^2,\mu^2)  =  B^\prime(p^2,\Lambda^2) \,,
\end{equation}
with $\lim_{\Lambda\to \infty} B^\prime(p^2,\Lambda^2) <
\infty$.\footnote{This is a model-independent statement; i.e., it is true in
any study that preserves at least the one-loop renormalisation group
behaviour of QCD.}  In terms of the renormalised current-quark mass the
existence of DCSB means that, in the chiral limit, $M(\mu^2) \sim {\rm
O}(1/\mu^2)$, up to $\ln\mu^2$-corrections.

In Fig.~\ref{spplot} we present the renormalised dressed-quark mass
function, $M(p^2)$, obtained by solving Eq.~(\ref{dsemod}) using the
parameters in Eq.~(\ref{params}), and in the chiral limit.  [Recall that
$\mu= 19\,$GeV, which is large enough to be in the perturbative domain.]

It is clear from this figure that the light-quark mass function is
characterised by a significant infrared enhancement, a direct result of that
in the effective coupling, ${\cal G}(k^2)$.  Introducing the Euclidean
constituent-quark mass, $M^E$, as the solution of $p^2=M^2(p^2)$, the ratio:
$M^E_f/ m_f(\mu)$, where $f$ labels the quark flavour and
$m_f(\mu)$ is given in Eq.~(\ref{params}), is a single, indicative and
quantitative measure of the nonperturbative effects of gluon-dressing on the
quark propagator.  We find
\begin{equation}
\label{cqm}
\begin{array}{llc}
                        &  M^E  & \displaystyle \frac{M^E}
                                                {m_f(\mu)} \\
\mbox{chiral limit}      &  0.55\, {\rm GeV} &     \infty        \\
u/d                     &  0.56          & 150        \\
s                       &  0.70           & 8.5        
\end{array}
\end{equation}
which clearly indicates the magnitude of this effect for light-quarks.  The
ratio $M^E/ m_f(\mu)$ takes a value of O$(1)$ for
heavy-quarks~\cite{ivanov} because the current-quark mass is much larger than
the mass-scale characterising the infrared enhancement in the effective
coupling, $\Lambda_{\rm QCD}$.  This means that in the spacelike region the
momentum-dependence of the heavy-quark mass function is dominated by
perturbative effects.\footnote{Quark confinement entails that there is no
``pole-mass''~\cite{confinement}, which would be the solution of
$p^2+M^2(p^2)=0$.  Hence, this definition of $M^E$ is arbitrary; a factor of
$2$ is certainly unimportant with respect to the qualitative features that
this quantity characterises.}

In typical quark model calculations~\cite{simon} the ``constituent-quark''
masses are $M_{u/d}=0.33\,$GeV and $M_s=0.55\,$GeV.  These are within a
factor of $2$ of the values in Eq.~(\ref{cqm}); and are in the ratio
$M_{u/d}/M_s = 0.60$.  From Eq.(\ref{cqm}) we find $M^E_{u/d}/M^E_s = 0.80$.
The comparison of numerous DSE studies makes it clear that this
correspondence between $M_f$ and $M^E_f$ is robust.  It provides a
qualitative understanding of the nature of the ``constituent-quark'' mass;
i.e., it is a quantitative measure of the nonperturbative modification of
quark propagation characteristics by gluon dressing.  Its magnitude is a
signal of the enhancement of the quark-quark interaction in the infrared.

The qualitative difference between the behaviour of $M(p^2)$ in the chiral
limit and in the presence of explicit chiral symmetry breaking is manifest in
Fig.~\ref{spplot}.  In the presence of explicit chiral symmetry breaking
Eq.~(\ref{masanom}) describes the form of $M(p^2)$ for $p^2 > {\rm O}(1\,{\rm
GeV}^2)$.  In the chiral limit, however, the ultraviolet behaviour is given
by
\begin{equation}
\label{Mchiral}
M(p^2) \stackrel{{\rm large}-p^2}{=}\,
\frac{2\pi^2\gamma_m}{3}\,\frac{\left(-\,\langle \bar q q \rangle^0\right)}
           {p^2
        \left(\case{1}{2}\ln\left[\frac{p^2}{\Lambda_{\rm QCD}^2}\right]
        \right)^{1-\gamma_m}}\,,
\end{equation}
where $\langle \bar q q \rangle^0$ is the renormalisation-point-independent
vacuum quark condensate.\footnote{The momentum dependence of this result is
characteristic of the QCD renormalisation group at one-loop~\cite{hdp76} and
demonstrates that the truncation we employ preserves this feature.}
Analysing our chiral limit solution we find
\begin{equation}
\label{qbqM0}
-\,\langle \bar q q \rangle^0 = (0.227\,{\rm GeV})^3\,.
\end{equation}
This is a reliable means of determining $\langle \bar q q \rangle^0$ because
corrections to Eq.~(\ref{Mchiral}) are suppressed by powers of $\Lambda_{\rm
QCD}^2/\mu^2$.

Equation~(\ref{cbqbq}) defines the renormalisation-point-dependent vacuum
quark condensate
\begin{equation}
\label{qbq19}
\left.-\,\langle \bar q q \rangle_\mu^0 \right|_{\mu=19\,{\rm GeV}}\doteq  
\left(\lim_{\Lambda\to\infty}
\left. Z_4(\mu,\Lambda)\, N_c \int^\Lambda_q\,{\rm tr}_{\rm Dirac}
        \left[ S_{\hat m =0}(q) \right]\right)\right|_{\mu=19\,{\rm GeV}}\,
= (0.275\,{\rm GeV})^3\,.
\end{equation}
We have established explicitly that $ m(\mu)\,\langle \bar q q
\rangle_\mu^0 =\,$constant, independent of $\mu$ with the value depending on the
quark flavour, and hence
\begin{equation}
\label{rgiprod}
m(\mu)\,\langle \bar q q \rangle_\mu^0 \doteq  \hat m \,\langle \bar q q
\rangle^0\,,
\end{equation}
which unambiguously defines the renormalisation-point-independent
current-quark masses.   From this and Eqs.~(\ref{params}), (\ref{qbqM0}) and
(\ref{qbq19}) we extract the values of these masses appropriate to our model
\begin{equation}
\label{rgimass}
\hat m_{u/d} = 6.60 \, {\rm MeV} \,,\;
\hat m_s = 147\, {\rm MeV}\,.
\end{equation}
Using Eq.~(\ref{masanom}) these values yield $m_{u/d}(\mu)= 3.2\,$MeV and
$m_s(\mu)= 72\,$MeV, which are within $\sim 10$\% of our actual values in
Eq.~(\ref{params}).  This indicates that higher-loop corrections to the
one-loop formulae, which are present in the solution of the integral equation
as made evident by $A(p^2,\mu^2)\not\equiv 1$, provide contributions of
$<10$\% at $p^2 = \mu^2$.  These contributions decrease with increasing
$p^2$.\footnote{Our model for the kernel of the quark-DSE is not constructed to
preserve the two-loop, perturbative behaviour of QCD.  Hence a direct
comparison at this level is not meaningful.}

From the renormalisation-point-invariant product in Eq.~(\ref{rgiprod}) we
obtain
\begin{equation}
\label{qbq1}
\left.-\,\langle \bar q q
\rangle_\mu^0\right|_{\mu=1\,{\rm GeV}}
\doteq \left(\ln\left[1/\Lambda_{\rm QCD}\right]\right)^{\gamma_m}
\, \langle \bar q q \rangle^0
= (0.241\,{\rm GeV})^3\,.
\end{equation}
This result can be compared directly with the value of the quark condensate
employed in contemporary phenomenological studies~\cite{derek}: $ (0.236\pm
0.008\,{\rm GeV})^3$.  We note that increasing $\omega \to 1.5\,\omega$ in
${\cal G}(k^2)$ increases the calculated value in Eq.~(\ref{qbq1}) by $\sim
10$\%.  Obtaining broad agreement with the contemporary phenomenological
value of $\langle \bar q q \rangle_{\mu=1\,{\rm GeV}}^0$ was a means we
employed to constrain the value of this parameter. However, we made no
attempt to fine-tune $\omega$ nor thereby our calculated value of $\langle
\bar q q \rangle_{\mu=1\,{\rm GeV}}^0$.

In conjunction with Eq.~(\ref{qbq1}) we define $m_f^{1\,{\rm GeV}}$ via
Eq.~(\ref{masanom}) using Eq.~(\ref{rgimass})
\begin{equation}
m_{u/d}^{1\,{\rm GeV}} = 5.5\,{\rm MeV}\,,\;
m_{s}^{1\,{\rm GeV}} = 130\,{\rm MeV}\,.
\end{equation}
These values differ slightly from those discussed in Sec.~\ref{secwmt} for
the reasons described above.  It is now clear, from Eq.~(\ref{rgiprod}), that
lower values of the current-quark masses, as canvassed in Ref.~\cite{kim97},
are admissible in our phenomenological study only via an increase in $\langle
\bar q q \rangle_{\mu=1\,{\rm GeV}}^0$.

After this discussion of the vacuum quark condensate it is now
straightforward to determine the accuracy of Eqs.~(\ref{gmorepi}) and
(\ref{gmoreKp}).  Using experimental values on the left-hand-side, we find:
\begin{eqnarray}
\label{picf}
(0.0924 \times 0.1385)^2 = (0.113\,{\rm GeV})^4 & \;{\rm cf.} \;& 
(0.111\,{\rm GeV})^4 = 2\times 0.0055 \times 0.24^3 \\
\label{Kcf}
(0.113 \times 0.495)^2 = (0.237\,{\rm GeV})^4& \;{\rm cf.} \;& 
(0.206\,{\rm GeV})^4 = (0.0055 + 0.13)\times 0.24^3\,,
\end{eqnarray}
which indicates that O$(\hat m^2)$-corrections begin to become important at
current-quark masses near that of the $s$-quark.  We emphasise that we did
not use these equations in fitting the current-quark masses but, as described
in Sec.~\ref{bsesoln}, solved the model BSE, Eq.~(\ref{bsemod}), using the
dressed-quark propagators obtained as a solution of Eq.~(\ref{dsemod}).  In
this procedure, changes in $\langle \bar q q \rangle^0$ effected by modifying
the model parameters are compensated for by changes in $\hat m_{u/d}$ and
$\hat m_s$.  Therefore the comparisons in Eqs.~(\ref{picf}) and (\ref{Kcf}),
which involve a product of these quantities, remain meaningful and the
conclusion remains valid.\footnote{Using our calculated values of $f_\pi$,
$m_\pi$, $f_K$ and $m_K$, Tables~\ref{respi} and \ref{resK}, the only change
is in Eq.~(\ref{Kcf}), where $0.237\to 0.233$.}

\subsection{Solution of the BSE}
\label{bsesoln}
To solve Eq.~(\ref{bsemod}) we modify it by introducing an eigenvalue,
$\lambda(P^2)$,
\begin{equation}
\label{bsemodlam}
\Gamma_H(k;P) + \lambda(P^2)\,\int^\Lambda_q
{\cal G}((k-q)^2)\, D_{\mu\nu}^{\rm free}(k-q)
 \frac{\lambda^a}{2}\gamma_\mu {\cal S}(q_+)\Gamma_H(q;P){\cal S}(q_-)
\frac{\lambda^a}{2}\gamma_\nu = 0\,,
\end{equation}
which yields an equation that has a solution $\forall\,P^2$, characterised by
the value of $\lambda(P^2)$.  The original problem is solved when that $P^2$
is found for which $\lambda(P^2)=1$, which will occur at $P^2<0$ in our
metric.

For $P^2=-{\cal P}^2 <0$, solving Eq.~(\ref{bsemodlam}) requires the
dressed-quark propagator ${\cal S}(q_+)$ on the parabola: $4\xi^2 {\cal
P}^2\Re(q_+^2)= \Im(q_+^2)^2 - 4(\xi^2 {\cal P}^2)^2 $, in the
complex-$q_+^2$ plane, and ${\cal S}(q_-)$ on the parabola: $4 (1-\xi)^2
{\cal P}^2\Re(q_-^2)= \Im(q_-^2)^2 -4 ((1-\xi)^2 {\cal P}^2)^2$, in the
complex-$q_-^2$ plane.  For complex arguments in the dressed-quark propagator
the quark-DSE requires the effective coupling at complex values of its
argument.  Equation~(\ref{bsemodlam}), however, still only requires ${\cal
G}(k^2)$ on the real-$k^2>0$ axis.  The specification of ${\cal G}(k^2)$ is
the primary element in the definition of the model and our Ansatz is
motivated by studies that are restricted to real-$k^2>0$.  In employing this
Ansatz at complex values of its arguments we are exploring an unconstrained
domain.  Solving numerically for ${\cal S}(p)$ in the complex-$p^2$ plane is
straightforward.  However, complex-conjugate branch points in the [confining]
solution introduce numerical complications in solving the BSE for bound
states containing a single heavy constituent.\footnote{This branch-point-pair
is present because of the infrared enhancement in our Ansatz for the
effective coupling but may be an artefact of the rainbow approximation,
Eq.~(\ref{rainbow}).  Ref.~\cite{brw92} demonstrates that dressing the
vertex, consistent with the constraints of Refs.~\cite{qedwti}, can
significantly affect the analytic properties of ${\cal S}(p)$ while
maintaining the essence of quark confinement~\cite{confinement}; i.e., that
${\cal S}(p)$ not have a Lehmann representation.}

The general form of the solution of the BSE, Eq.~(\ref{bsemod}), is given in
Eq.~(\ref{genpibsa}) where the scalar functions depend on the variables $k^2$
and $k\cdot P$, and are labelled by the eigenvalue $P^2$.  From this it is
clear that the integrand in Eq.~(\ref{bsemodlam}) depends on the scalars:
$k^2$, $k\cdot q$, $q^2$, $q\cdot P$ and $P^2$, which takes a fixed-value at
the solution; i.e., at each value of $P^2$ the kernel is a function of four,
independent variables.  Solving Eq.~(\ref{bsemodlam}) can therefore require
large-scale computing resources, especially since there are four, independent
scalar functions in the general form of the solution.

We employed two different techniques in solving Eq.~(\ref{bsemodlam}).  In
our primary procedure, (A), we treated the scalar functions directly as
dependent on two, independent variables: $E(k^2,k\cdot P; P^2)$, etc., which
requires straightforward, multidimensional integration at every iteration.
Storing the multidimensional kernel requires a large amount of computer
memory but the iteration proceeds quickly.

As an adjunct, and because we wish to elucidate qualitative and quantitative
effects related to the $k\cdot P$-dependence of the scalar functions in the
BSA, we employed a Chebyshev decomposition procedure, (B).  To implement this
we write
\begin{equation}
\label{chebexp}
E(k^2,k\cdot P;P^2) \approx \sum_{i=0}^{N_{\rm max}}
 \,^i\!E(k^2;P^2)\,U_i(\cos\beta)\,,
\end{equation}
with similar expansions for $F$, \underline{$\hat G\doteq k\cdot P\,G$} and
$H$, where $k\cdot P \doteq \cos\beta\sqrt{k^2 P^2}$ and
$\{U_i(x);i=0,\ldots,\infty\}$ are Chebyshev polynomials of the second kind
orthonormalised according to:
\begin{equation}
\frac{2}{\pi}\int_{-1}^1\,dx\,\sqrt{1-x^2}\, U_i(x) U_j(x) = \delta_{ij}\,.
\end{equation}
Substituting this expansion into Eq.~(\ref{bsemodlam}) allows all-but-one of
the integrals to be evaluated before beginning the iteration.  One then
solves for the Chebyshev moments, $^i\!E(k^2;P^2)$.  This procedure requires
a large amount of time to set up the kernel but does not require large
amounts of computer memory.\footnote{Equation~(\ref{chebexp}) is only an
identity in the limit $N_{\rm max}\to \infty$ but, in the present example, an
accurate representation of the solution is obtained with $N_{\rm max} =
1\;{\rm or} \;2$, which is consistent with the observations of
Ref.~\cite{tjon}.}

In order to fit the parameter $D$ in ${\cal G}(k^2)$ and the current-quark
mass $ m_{u/d}(\mu)$ we: 1) chose values for these parameters; 2) solved the
quark DSE for $S_{u/d}(p)$; 3) used ${\cal G}(k^2)$ and the calculated form
of $S_{u/d}(p)$ to solve the pion BSE for the mass and BSA,
$\Gamma_\pi(k;P)$, using procedure (A); and 4) used the calculated forms of
$S_{u/d}(p)$ and $\Gamma_\pi(k;P)$ to calculate $f_\pi$ from
Eq.~(\ref{caint}).  We repeated this procedure until satisfactory values of
$m_\pi$ and $f_\pi$ were obtained.  Having thus fixed $D$ we repeated the steps
for the $K$-meson, varying $ m_s(\mu)$ only, in order to obtain the best
possible values of $m_K$ and $f_K$.  This led to the parameter values quoted in
Eq.~(\ref{params}) and the results listed in row one of Tables~\ref{respi} and
\ref{resK}.  This does not represent an exhaustive search of the available
parameter space but is sufficient for our purposes.

\subsubsection{Discussion of the BSE solution}
\label{discussbse}
From Tables~\ref{respi} and \ref{resK}, and Eqs.~(\ref{params}),
(\ref{gmorqbqM}), (\ref{qbqM}) and (\ref{qbq1}), it is straightforward to
calculate
\begin{equation}
\begin{array}{ccc}
 -\langle \bar q q \rangle^\pi_{\mu = 1\,{\rm GeV}} &  
-\langle \bar q q\rangle^K_{\mu = 1\,{\rm GeV}} & 
 -\langle\bar q q \rangle^{s\bar s}_{\mu = 1\,{\rm GeV}} \\
& &\\
(0.245 \, {\rm GeV})^3 &
(0.284 \, {\rm GeV})^3 &
(0.317 \, {\rm GeV})^3 
\end{array}
\end{equation}
showing that, for light pseudoscalars, the ``in-meson condensate'' that we
have defined increases with increasing bound state mass; as does the leptonic
decay constant, $f_H$.\footnote{$(-\langle \bar q q \rangle^H_{\mu })/f_H$ is
the residue of the bound state pole in the pseudoscalar vertex, just as $f_H$
is the residue of the bound state pole in the axial-vector vertex.  As
expected, $\langle\bar q q\rangle^\pi_{\mu=1\,{\rm GeV}}\approx
\left.\langle\bar q q\rangle^0_\mu\right|_{\mu=1\,{\rm GeV}}$.}

In Tables~\ref{respi} and \ref{resK} we list values of the dimensionless
ratio
\begin{equation}
{\cal R}_H \doteq \,-\,\frac{\langle \bar q q \rangle_\mu^H {\cal M}_H}
        {f_H^2 m_H^2}\,.
\end{equation}
A value of ${\cal R}_H=1$ means that Eq.~(\ref{gmora}) is satisfied and hence
so is the AV-WTI.\footnote{It illustrates that the pseudoscalar-meson pole in
the axial-vector vertex is related to the pseudoscalar-meson pole in the
pseudoscalar vertex in the manner we have elucidated.  A finite value in the
chiral limit emphasises that $m_H^2 \propto {\cal M}_H$ as ${\cal M}_H\to
0$.}  Looking at the tabulated values of ${\cal R}_H$ it is clear that the
scalar function $H$ is not quantitatively important, with the AV-WTI being
satisfied numerically with the retention of $E$, $F$ and $G$ in the
pseudoscalar meson BSA.  The values of ${\cal R}_H$, and the other tabulated
quantities, highlight the importance of $F$ and $\hat G$: $F$ is the most
important of these functions but $\hat G$ nevertheless provides a significant
contribution, particularly for bound states of unequal-mass constituents.  We
note that a poor value of $f_H$ is tied to a poor value of ${\cal R}_H$,
which emphasises the importance of preserving the AV-WTI and hence
Eq.~(\ref{npifpi}).  We have checked explicitly that our complete solutions
satisfy Eq.~(\ref{npifpi}).

The tables illustrate the rapid convergence of the Chebyshev decomposition
procedure, (B), with accurate solutions being obtained with the zeroth and
first moments for bound states of equal mass constituents.\footnote{We note
that for equal-mass constituents $\hat G = k\cdot P\, G$ is an odd function
of $k\cdot P$ and hence $^0\!\hat G\equiv 0$.  Therefore it first contributes
at O$(U_1)$.}  The same is true for bound states of unequal mass constituents
{\it provided} all Dirac amplitudes are retained in the solution.

One observes that when the $k\cdot P$-dependence of the scalar functions in
the meson Bethe-Salpeter amplitude is included, physical observables are {\it
independent} of the momentum partitioning parameter, $\eta_P$.  These
calculations elucidate the manner in which this necessary requirement in
covariant bound state studies is realised: the bound state amplitude depends
on $\eta_P$ in just that fashion which ensures physical quantities do not.
All the scalar functions in the BSA must be included to ensure this.

The tables also illustrate clearly the effect of the current-quark mass via a
comparison of the chiral limit pion with the physical pion, the kaon and a
fictitious, pseudoscalar, $s\bar s$ bound state.  One observes that $f_H$ is
weakly sensitive to increasing the current-quark mass; for example, we obtain
$f^0/f_\pi=0.97$, which is consistent with expectations based on effective
chiral Lagrangians.  However, $m_H^2$ rises quickly, before becoming
sensitive to effects nonlinear in the current-quark mass.\footnote{We note
that our calculated value of $m_{s\bar s}$ is within 3\% of that obtained for
this fictitious bound state in Ref.~\cite{jm93}.  However, the results in
Ref.~\cite{jm93} depend on $\eta_P$, which is an artefact of the solution
procedure adopted therein.  In that study a derivative-expansion and
extrapolation procedure was employed in order to avoid a direct solution for
the quark propagator functions $A(p^2)$ and $B(p^2)$ at complex values of
their arguments.  In addition, in most instances only the zeroth Chebyshev
moment was retained in the Chebyshev expansion of the scalar functions in the
Bethe-Salpeter {\it wave function}$\,$: $\chi_H(q;P) \doteq {\cal S}(q_+)
\Gamma_H(q;P) {\cal S}(q_-)$.}

We present the scalar functions in the BSA obtained as solutions of
Eq.~(\ref{bsemod}) in Figs.~\ref{figE}-\ref{figUV}, focusing on the zeroth
Chebyshev moment of each function, which is obtained via
\begin{equation}
^0\!E_H(k^2) \doteq 
        \frac{2}{\pi}\int_0^\pi\,d\beta\,\sin^2\beta\,U_0(\cos\beta)\,
                        E_H(k^2,k\cdot P;P^2)\,,
\end{equation}
and similarly for $F$, $G$ [$\hat G$ for the $K$-meson] and $H$.
Figure~\ref{figE} illustrates that the momentum-space width of $^0\!E_H(k^2)$
increases as the current-quark mass of the bound state constituents
increases; Fig.~\ref{figF}, that $^0\!F_H(k^2=0)$ decreases with increasing
current-quark mass but that $^0\!F_H(k^2)$ is still larger at
$k^2>0.5\,$GeV$^2$ for bound states of higher mass; Fig.~\ref{figG}, that
$^0\!G_H(k^2)$ [$^0\!\hat G_K(k^2)$] behaves similarly; and Fig.~\ref{figH},
that the same is true for $H_H(k;P)$ and that it is uniformly small in
magnitude thereby explaining its quantitative insignificance.  Where
comparison is possible, these observations agree qualitatively with
Refs.~\cite{jm93,sep97}.

In Fig.~\ref{figUV} we illustrate the large-$k^2$ behaviour of the scalar
functions in the pseudoscalar BSA.  The momentum dependence of $^0\!E_H(k^2)$
at large-$k^2$ is identical to that of the chiral-limit quark mass function,
$M(p^2)$, in Eq.~(\ref{Mchiral})~\cite{m90}, and characterises the form of
the quark-quark interaction in the ultraviolet.  Figure~\ref{figUV}
elucidates that this is also true of $^0\!F_H(k^2)$, $k^2\,^0\!G_H(k^2)$
[$k^2\,^0\!\hat G_K(k^2)$ for the $K$-meson] and $k^2\,^0\!H_H(k^2)$.  Each
of these functions reaches its ultraviolet limit by $k^2 \simeq 10\,$GeV$^2$,
which is very-much-less-than the renormalisation point, $\mu^2=361\,$GeV$^2$.

In order to verify Eqs.~(\ref{bwti})-(\ref{gwti}) it is necessary to consider
the inhomogeneous axial-vector vertex equation, Eq.~(\ref{genave}), in our
truncation of the DSEs, which is
\begin{equation}
\label{ihbseav}
\Gamma_{5\mu}^H(k;P) = 
Z_2 \gamma_5\gamma_\mu \frac{T^H}{2}-
\int^\Lambda_q
{\cal G}((k-q)^2)\, D_{\mu\nu}^{\rm free}(k-q)
 \frac{\lambda^a}{2}\gamma_\mu {\cal S}(q_+)\Gamma_{5\mu}^H(q;P){\cal S}(q_-)
\frac{\lambda^a}{2}\gamma_\nu\,.
\end{equation}
From the homogeneous BSE one already has the equations satisfied by
$E_H(k;0)$, $F_H(k;0)$, $G_H(k;0)$ and $H_H(k;0)$.  To proceed, one
substitutes Eq.~(\ref{truavv}) for $\Gamma_{5\mu}^H(k;P)$ in
Eq.~(\ref{ihbseav}).  Using the coupled equations for $E_H(k;0)$, etc., one
can identify and eliminate each of the pole terms associated with
pseudoscalar bound state.  [It is the fact that the homogeneous BSE is linear
in the BSA that allows this.]  This yields a system of coupled equations for
$F_R(k;0)$, $G_R(k;0)$ and $H_R(k;0)$, which can be solved without
complication.  [The factor of $Z_2$ automatically ensures that
$F_R(k^2=\mu^2;P=0)=1$.]  We illustrate the realisation of the first two
identities, Eqs.~(\ref{bwti}) and (\ref{fwti}), in Fig.~\ref{wtiplot}.  The
remaining two identities, Eqs.~(\protect\ref{rgwti}) and (\protect\ref{gwti}),
are realised in a similar fashion.  

\section{Summary and Conclusion}
With renormalisation considered explicitly, we have studied the Dyson
Schwinger equation [DSE] for the dressed-quark propagator, $S(p)$; the
homogeneous, pseudoscalar meson Bethe-Salpeter equation [BSE], which provides
the meson's Bethe-Salpeter amplitude, $\Gamma(k;P)$, with $P$ the total
momentum of the bound state; and the inhomogeneous BSEs for the
fully-amputated axial-vector and pseudoscalar vertices, $\Gamma_{5\mu}(k;P)$
and $\Gamma_{5}(k;P)$, respectively.  Independent of assumptions about the
form of the quark-quark scattering kernel, $K(q,k;P)$, we have elucidated the
manner in which the axial-vector Ward-Takahashi identity [AV-WTI] correlates
these Schwinger functions and provides important and phenomenologically
useful constraints.

We demonstrated that the axial-vector vertex contains a pseudoscalar, bound
state pole contribution whose residue is the meson's leptonic decay
constant, Eq.~(\ref{caint}).  The pseudoscalar vertex also contains such a
pole term but in this case the residue is related to an ``in-meson'' quark
condensate, Eq.~(\ref{rH}), which is equal to the vacuum quark condensate in
the chiral limit, Eq.~(\ref{cbqbq}).  The AV-WTI necessarily entails an
identity between these residues, Eq.~(\ref{gmora}), which is valid {\it
independent} of the current-quark mass of the bound state constituents.  The
expression commonly known as the Gell-Mann--Oakes--Renner relation is a
corollary of this identity.  The AV-WTI also places constraints on the form
of the pseudoscalar meson Bethe-Salpeter amplitude [BSA], which necessarily
involves quantitatively important terms proportional to $\gamma_5\gamma\cdot
P$ and $\gamma\cdot k \,k\cdot P$ that have hitherto been neglected in
phenomenological studies of spectra, and production and scattering processes
in QCD.  In the chiral limit these constraints take a very simple form:
Eqs.~(\ref{bwti})-(\ref{gwti}).

In Sec.~\ref{discussbse} we illustrated these identities and constraints in
numerical studies using a rainbow-ladder truncation for $K(q,k;P)$ motivated
by the Abelian approximation, Secs.~\ref{modeldse} and \ref{secmodbse}, and
defined by an Ansatz for the dressed-quark-quark interaction,
Fig.\ref{gkplot}.  The model thus defined preserves the one-loop
renormalisation-group behaviour of QCD, as is clear in our numerical
solutions.  This aspect of our study facilitated an elucidation of the
dominant, ultraviolet [large-$k^2$] behaviour of the scalar functions in the
pseudoscalar meson BSA, Fig.~\ref{figUV}.  Employing these results in a
calculation of the electromagnetic pion form factor yields $Q^2 \,
F_\pi(Q^2)=\,$constant, up to $\ln Q^2$-corrections; a direct consequence of
the model-independent result: $^0\!F_H(k^2) \sim k^2\,^0\!\hat G_K(k^2) \sim
1/[k^2(\ln k/\Lambda_{\rm QCD})^{1-\gamma_m}]$~\cite{MRnew}.  The
pseudoscalar amplitude $E_\pi(k;P)$ does not contribute to the asymptotic
form of $F_\pi(Q^2)$.\footnote{We note that the calculation of the elastic
pion form factor involves two pion Bethe-Salpeter amplitudes, which is
qualitatively different to the electroproduction studies of
Ref.~\cite{pich97}, which involve only one meson Bethe-Salpeter amplitude.
We therefore expect that the conclusions of Ref.~\cite{pich97} will not be
qualitatively sensitive to the omission of all vector meson Dirac amplitudes
other than $\gamma_\mu$ since, in that calculation, the Bethe-Salpeter
amplitude acts only to restrict the support of the integrand and the
asymptotic behaviour of the cross sections is determined by the behaviour of
the quark propagator.}

In the course of these numerical studies we explored the feasibility of two
methods for solving the homogeneous Bethe-Salpeter equation with
dressed-quark propagators, determined numerically, and without a
three-dimensional reduction: (A) treating the problem directly as a
multidimensional integral equation; or (B) employing a Chebyshev expansion of
the scalar functions in the BSA in order to obtain a system of
one-dimensional integral equations.  We favoured (A) because it provides a
simpler numerical procedure, the programme runs quickly and computer memory
usage was not a consideration.  Nevertheless, we observed that the Chebyshev
expansion converged quickly, with at most two moments being necessary to
reproduce the solution obtained using (A), Sec.~\ref{discussbse}.
Importantly, we saw that retaining the $k\cdot P$-dependence of the scalar
functions in the meson BSA is all that is necessary to ensure that physical
observables are independent of the momentum partitioning parameter that
appears in the definition of the relative momentum, which is arbitrary in
covariant studies.

We have demonstrated that the Goldstone boson character of flavour-nonsinglet,
pseudoscalar mesons is no impediment to their description as bound states in
QCD.  It is only important that, in developing the bound state equations, the
axial-vector Ward-Takahashi identity be preserved explicitly.  This identity
necessarily entails relations between the $n$-point Schwinger functions
(propagators and vertices) relevant to the bound state equation; relations
which provide for the manifestation of Goldstone's theorem.  In taking
account of this, one can in principle construct a single kernel for the
Bethe-Salpeter equation that will provide a uniformly good, qualitative and
quantitative description of the properties of all mesons.

\acknowledgements We acknowledge useful conversations and correspondence with
A. Bender, M. C. Birse, F. T. Hawes, J. A. McGovern, P. C. Tandy and
A.~G.~Williams.  CDR is grateful to the Department of Physics and
Mathematical Physics at the University of Adelaide for their hospitality and
support during a term as a Distinguished Visiting Scholar in which some of
this work was conducted.  This work was funded by the US Department of
Energy, Nuclear Physics Division, under contract number W-31-109-ENG-38; and
benefited from the resources of the National Energy Research Scientific
Computing Center.


\begin{figure}
\centering{\
\epsfig{figure=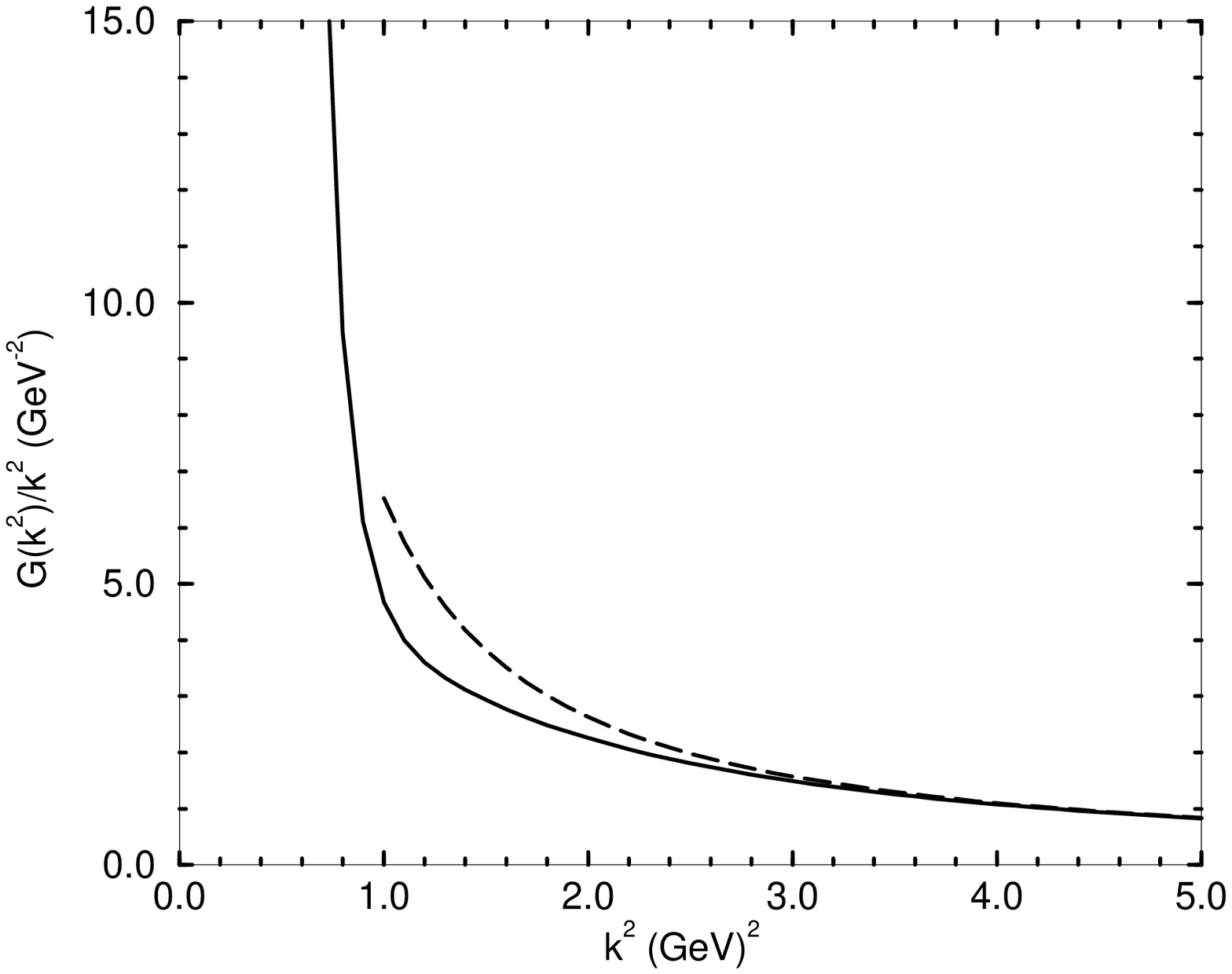,height=13.0cm}}
\caption{A comparison of ${\cal G}(k^2)/k^2$ in Eq.~(\protect\ref{gk2})
obtained using the best-fit parameters of Eq.~(\protect\ref{params}) (solid
line), with $4\pi\alpha(k^2)/k^2$ in Eq.~(\protect\ref{alphaq2}) (dashed line).
The obvious infrared enhancement is qualitatively and semiquantitatively in
agreement with that inferred in the gluon DSE studies of
Ref.~\protect\cite{bp89}.
\label{gkplot}}
\end{figure}
\begin{figure}
\centering{\
\epsfig{figure=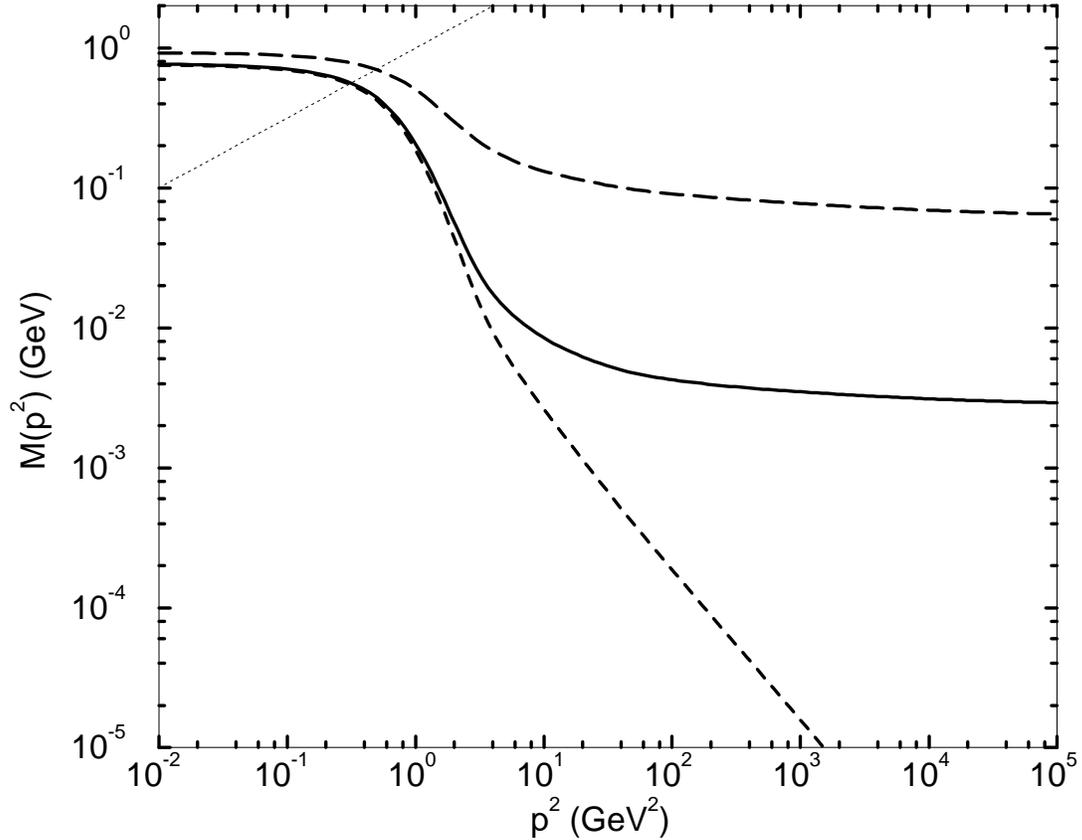,height=13.0cm}}
\caption{The renormalised dressed-quark mass function, $M(p^2)$, obtained by
solving Eq.~(\protect\ref{dsemod}) using the parameters in
Eq.~(\protect\ref{params}): $u/d$-quark (solid line); $s$-quark (long-dashed
line); and chiral limit (dashed line).  The renormalisation point is
$\mu=19\,$GeV.  The intersection of the line $M^2(p)=p^2$ (dotted line)
with each curve defines the Euclidean constituent-quark mass, $M^E$.
\label{spplot}}
\end{figure}
\begin{figure}
\centering{\
\epsfig{figure=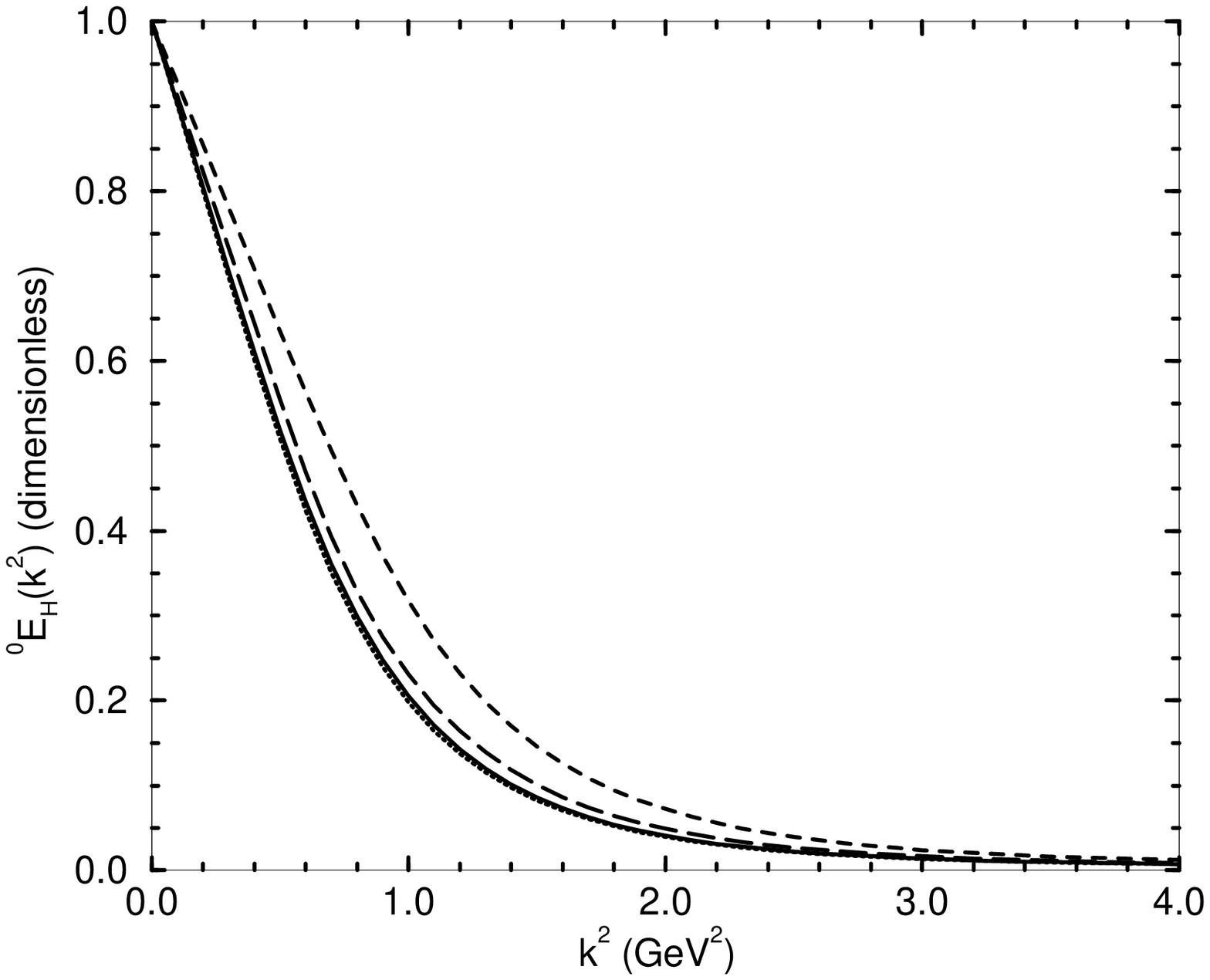,height=13.0cm}}
\caption{The zeroth Chebyshev moment of $E_H(k;P)$: chiral limit (dotted
line); $\pi$-meson (solid line); $K$-meson (long-dashed line); and fictitious,
$s\bar s$ bound state (dashed line).  $\eta_P=\case{1}{2}$ in each case.  For
ease of comparison the BSAs are all rescaled so that $^0\!E_H(k^2=0)=1$.
\label{figE}}
\end{figure}
\begin{figure}
\centering{\
\epsfig{figure=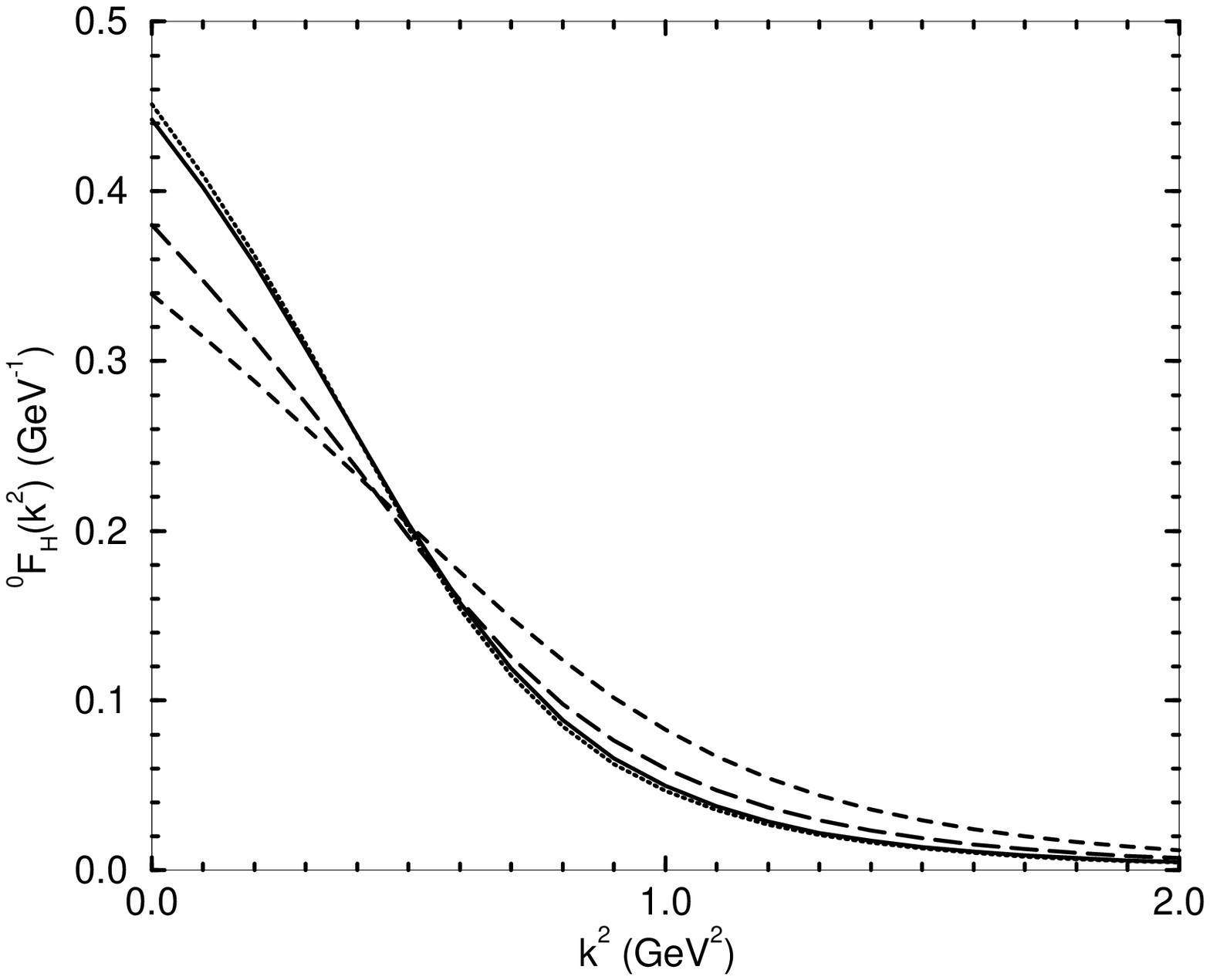,height=13.0cm}}
\caption{The zeroth Chebyshev moment of $F_H(k;P)$: chiral limit (dotted
line); $\pi$-meson (solid line); $K$-meson (long-dashed line); and fictitious,
$s\bar s$ bound state (dashed line).  $\eta_P=\case{1}{2}$ in each case.  For
ease of comparison the BSAs are all rescaled so that $^0\!E_H(k^2=0)=1$.
\label{figF}}
\end{figure}
\begin{figure}
\centering{\
\epsfig{figure=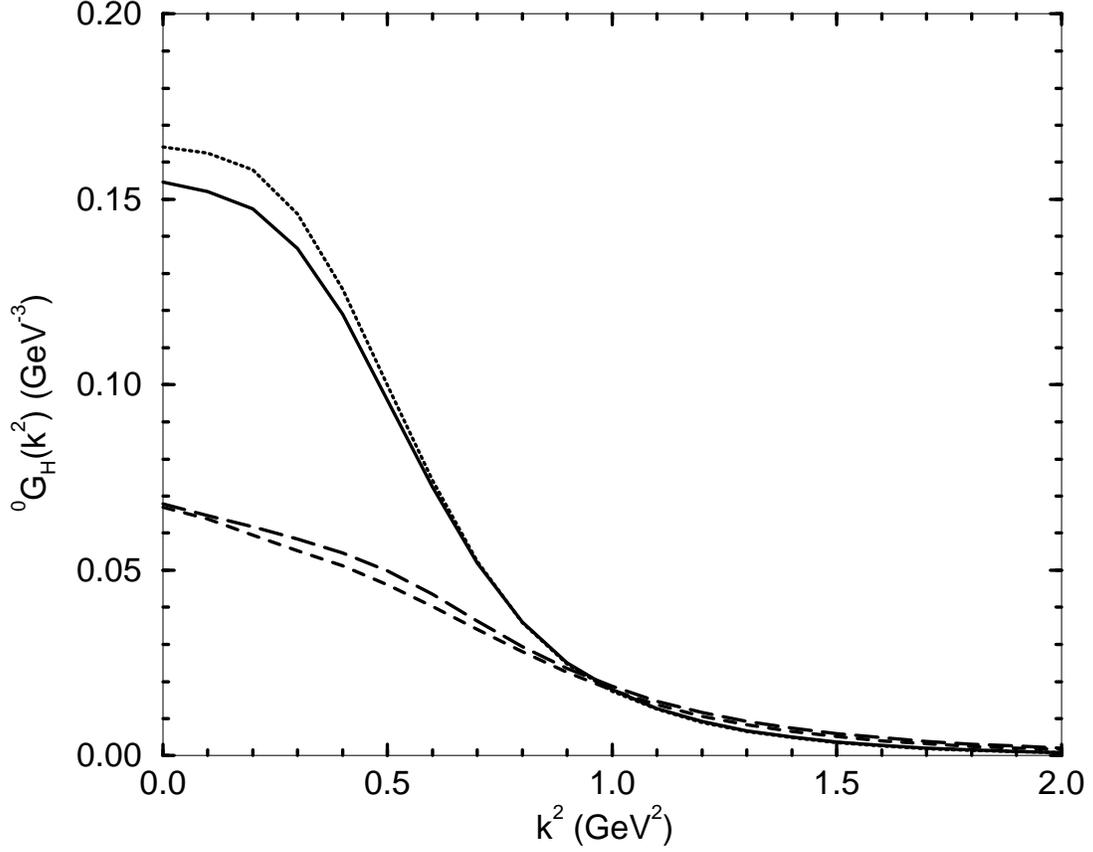,height=13.0cm}}
\caption{The zeroth Chebyshev moment of $G_H(k;P)$: chiral limit (dotted
line); $\pi$-meson (solid line); fictitious, $s\bar s$ bound state (dashed
line); and of $\hat G_H(k;P)\,({\rm GeV}^{-1})$ for the $K$-meson (long-dashed
line).  $\eta_P=\case{1}{2}$ in each case.  For ease of comparison the BSAs
are all rescaled so that $^0\!E_H(k^2=0)=1$.
\label{figG}}
\end{figure}
\begin{figure}
\centering{\
\epsfig{figure=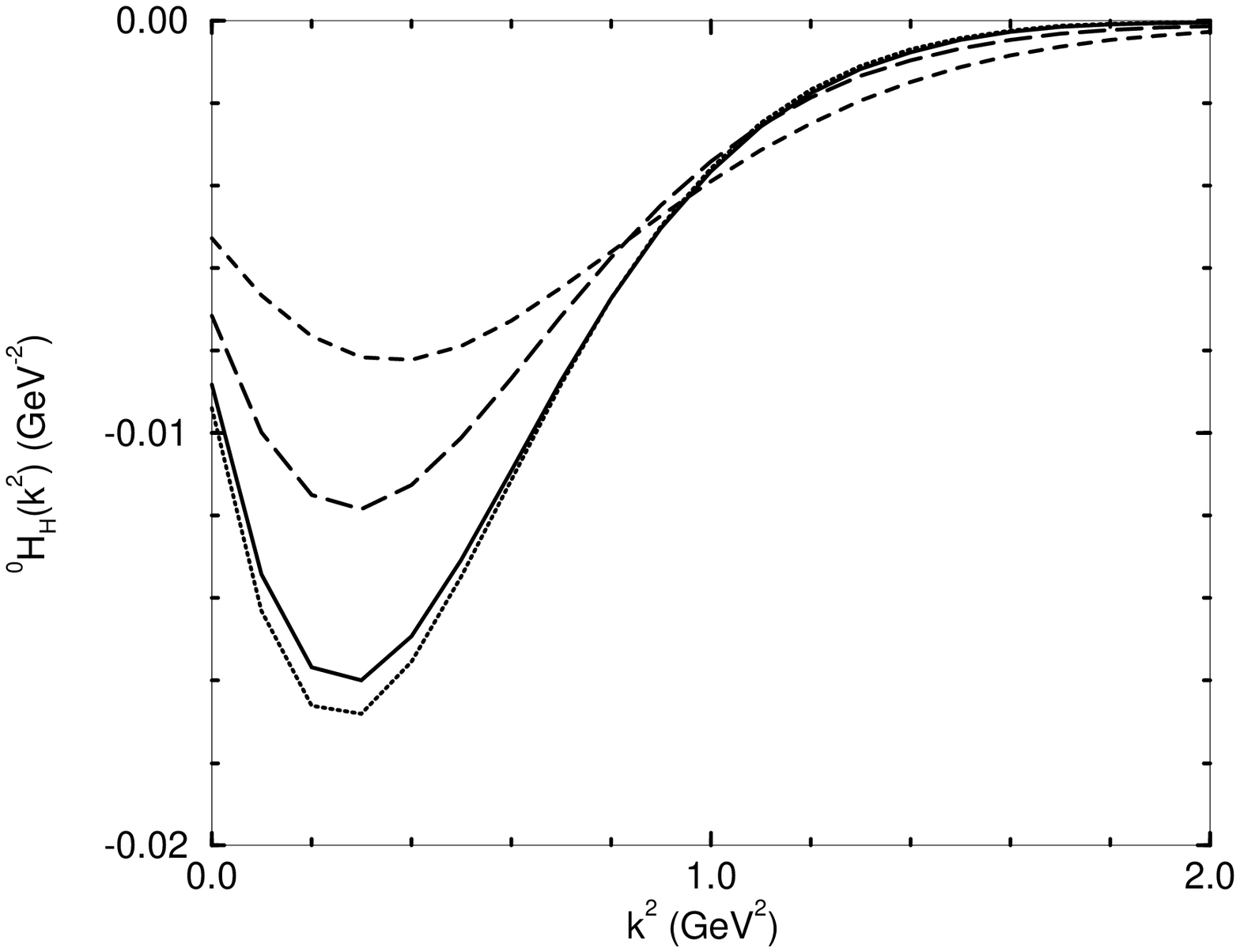,height=13.0cm}}
\caption{The zeroth Chebyshev moment of $H_H(k;P)$: chiral limit (dotted
line); $\pi$-meson (solid line); $K$-meson (long-dashed line); and fictitious,
$s\bar s$ bound state (dashed line).  $\eta_P=\case{1}{2}$ in each case.  For
ease of comparison the BSAs are all rescaled so that $^0\!E_H(k^2=0)=1$.
\label{figH}}
\end{figure}
\begin{figure}
\centering{\
\epsfig{figure=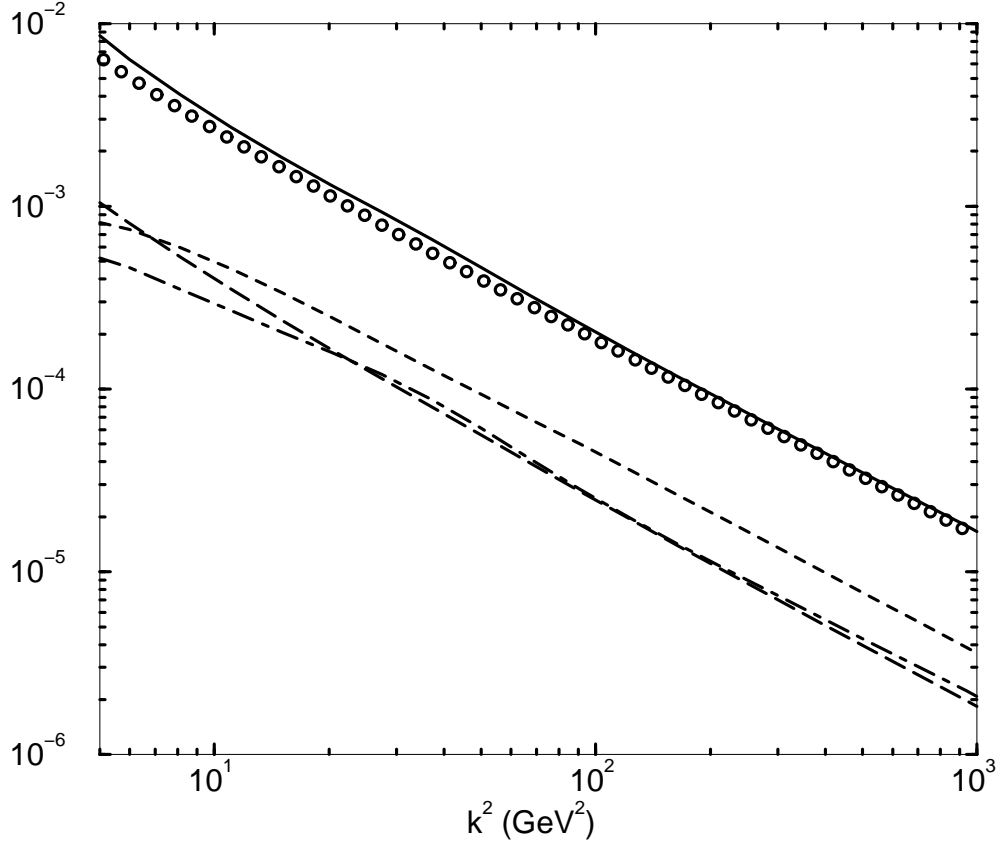,height=13.0cm}}
\caption{The asymptotic behaviour of the zeroth Chebyshev moments of the
functions in the $\pi$-meson BSA: $f_\pi\,^0\!E_\pi(k^2)$ (GeV, solid line);
$f_\pi\,^0\!F_\pi(k^2)$ (dimensionless, long-dashed line);
$k^2\,f_\pi\,^0\!G_\pi(k^2)$ (dimensionless, dashed line); and
$k^2\,f_\pi\,^0\!H_\pi(k^2)$ (GeV, dot-dashed line).  The momentum-dependence
is identical to that of the chiral-limit quark mass function, $M(p^2)$,
Eq.~(\protect\ref{Mchiral}) (GeV, circles: $\circ$).  For other pseudoscalar
mesons the momentum dependence of these functions is qualitatively the same,
although the normalising magnitude differs.
\label{figUV}}
\end{figure}
\begin{figure}
\centering{\
\epsfig{figure=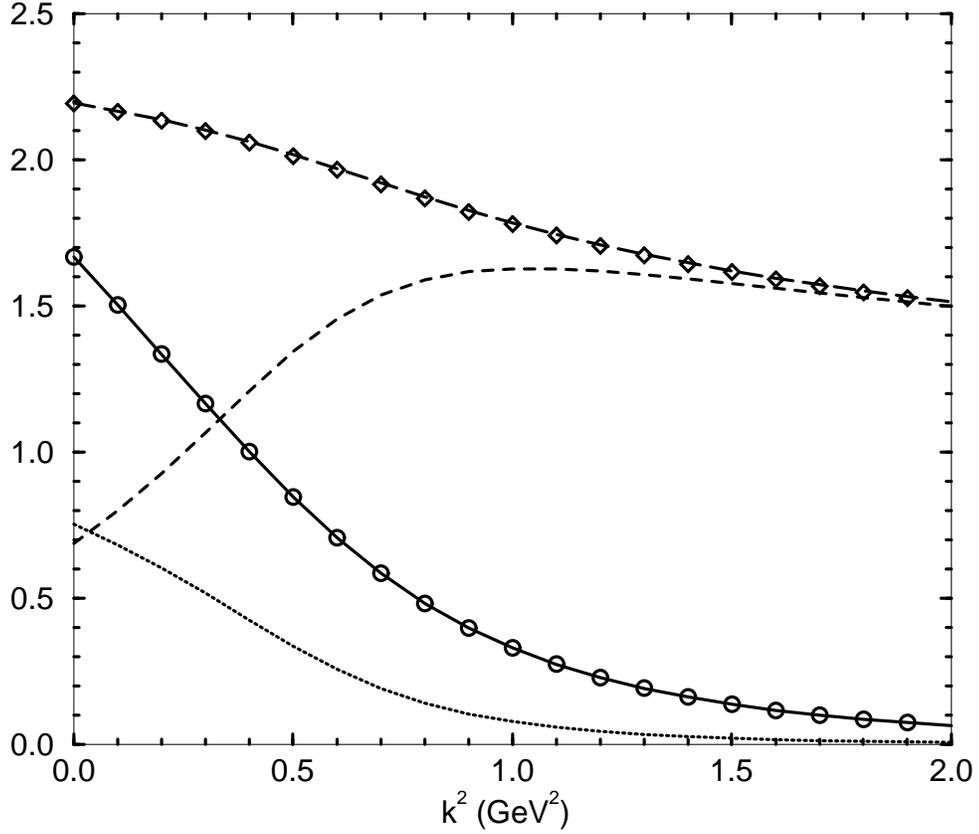,height=13.0cm}}
\caption{An illustration of the realisation in our model of the identities
Eqs.~(\protect\ref{bwti}) and (\protect\ref{fwti}), which are a necessary
consequence of preserving the axial vector Ward-Takahashi identity.  We plot
$f^0 E_H(k;0)$ (GeV, solid line); $F_R(k;0)$ (dimensionless, dashed line);
$f^0 F_H(k;0)$ (dimensionless, dotted line); and $ F_R(k;0) + 2 \, f_H
F_H(k;0) $ (long-dashed line).  In each curve the plotted points represent
the right-hand-side of these equations as obtained in the solution of the
chiral-limit quark DSE: $B(k^2)$ (GeV, $\circ$); $A(k^2)$ (dimensionless,
$\diamond$).
\label{wtiplot}}
\end{figure}
%
\begin{table}
\begin{tabular}{l|ddd|ddd|ddd}
All amplitudes & 
\multicolumn{3}{c|}{$\pi$} & 
\multicolumn{3}{c|}{chiral limit} & 
\multicolumn{3}{c}{$s\bar s$}\\
& $m_\pi$ & $f_\pi$ & ${\cal R}_H$ &
  $m_0$   & $f^0$   & ${\cal R}_H$ & 
$m_{s\bar s}$ & $f_{s\bar s}$ & ${\cal R}_H$ \\\hline
Method (A) &
0.1385 & 0.0924 & 1.01 & 0.0 & 0.0898 & 1.00 & 0.685 & 0.129 & 1.00 \\
$U_0$ only & 
0.136 & 0.0999 & 0.95 & 0.0 & 0.0972 & 0.94 & 0.675 & 0.137 & 0.95 \\
$U_0$ and $U_1$ &
0.1385 & 0.0925 & 1.00 & 0.0 & 0.0898 & 1.00 & 0.685 & 0.129 & 1.00  
\\ \hline\hline
$E$ only & & & & & & & & & \\\hline
Method (A)
& 0.105 & 0.0667 & 1.82  & 0.0 & 0.0649 & 1.81  & 0.512 & 0.092 & 1.68\\
$U_0$ only & 
0.105 & 0.0667 &1.82 & 0.0 & 0.0649 & 1.81 & 0.513 & 0.092 & 1.69  
\\\hline\hline
$E$, $F$ & & & & & & & & & \\\hline
Method (A)
& 0.136 & 0.0992 & 0.95  & 0.0 & 0.0965 & 0.95  & 0.677 & 0.137 & 0.95\\
$U_0$ only & 
0.136 & 0.0992 & 0.95 & 0.0 & 0.0965 & 0.95 & 0.678 & 0.138 & 0.95 
\\\hline\hline
$E$, $F$, $\rule{0mm}{4mm}\hat G$ & & & & & & & & & \\\hline
Method (A)
& 0.140 & 0.0917 & 1.01  & 0.0 & 0.0891 & 1.00  & 0.688 & 0.128 & 1.01\\
$U_0$ only & 
0.136 & 0.0992 & 0.95 & 0.0 & 0.0965 & 0.95 & 0.678 & 0.138 & 0.95 \\
$U_0$ and $U_1$ & 
0.140 & 0.0917 & 1.01 & 0.0 & 0.0891 & 1.00 & 0.689 & 0.128 & 1.01  
\end{tabular}
\caption{Calculated values of the properties of light, pseudoscalar mesons
composed of a quark and antiquark of equal-mass.  The mass $(m_\pi^{\rm
exp}=0.1385)$ and decay constant ($f_\pi^{\rm exp}=0.0924$) are in GeV,
${\cal R}_H$ is dimensionless.  With the exception of the calculations that
retain only the zeroth Chebyshev moment, labelled by ``$U_0$ only'', the
results are independent of the momentum partitioning parameter, $\eta_P$.
\label{respi}}
\end{table}
\begin{table}
\begin{tabular}{l|ddd|ddd|ddd}
All amplitudes & 
\multicolumn{3}{c|}{$\eta_P=0.50$} & 
\multicolumn{3}{c|}{$\eta_P=0.25$} & 
\multicolumn{3}{c}{$\eta_P=0.00$}\\
& $m_K$ & $f_K$ & ${\cal R}_K$ &
  $m_K$ & $f_K$ & ${\cal R}_K$ & 
  $m_K$ & $f_K$ & ${\cal R}_K$ \\\hline
Method (A) &
0.497 & 0.109 & 1.01 & 0.497 & 0.109 & 1.01 & 0.497 & 0.109 & 1.01 \\
$U_0$ only & 
0.469 & 0.117 & 0.96 & 0.482 & 0.117 & 0.95 & 0.475 & 0.113 & 1.02 \\
$U_0$ and $U_1$ & 
0.500 & 0.111 & 1.00 & 0.497 & 0.109 & 1.01 & 0.498 & 0.110 & 1.00 \\
$U_0$, $U_1$ and $U_2$ & 
0.497 & 0.109 & 1.01 & 0.497 & 0.109 & 1.01 & 0.496 & 0.109 & 1.01 
\\\hline\hline
$E$ only & & & & & & & & & \\\hline
Method (A) &
0.430 & 0.079 & 1.55 & 0.430 & 0.079 & 1.55 & 0.429 & 0.076 & 1.55 \\
$U_0$ only & 
0.380 & 0.077 & 1.54 & 0.401 & 0.076 & 1.51 & 0.415 & 0.073 & 1.55 \\
$U_0$ and $U_1$ & 
0.439 & 0.089 & 1.52 & 0.430 & 0.078 & 1.55 & 0.431 & 0.076 & 1.57
\\
$U_0$, $U_1$ and $U_2$ & 
0.430 & 0.078 & 1.55 & 0.430 & 0.078 & 1.55 & 0.427 & 0.076 & 1.55 
\\\hline\hline
$E$, $F$ & & & & & & & & & \\\hline
Method (A)
& 0.587 & 0.17  & 0.79  & 0.557 & 0.14  & 0.86  & 0.533 & 0.11 & 0.94\\
$U_0$ only & 
0.505 & 0.12  & 0.82 & 0.518 & 0.11  & 0.86 & 0.512 & 0.11  & 0.96 \\
$U_0$ and $U_1$ & 
 -    &  -     &  -   & 0.556 & 0.14  & 0.86 & 0.537 & 0.12  & 0.94  
\\
$U_0$, $U_1$ and $U_2$ & 
0.583 & 0.16  & 0.79 & 0.557 & 0.14  & 0.86 & 0.532 & 0.12  & 0.93 \\\hline\hline
$E$, $F$, $\rule{0mm}{4mm}\hat G$ & & & & & & & & & \\\hline
Method (A)
& 0.500 & 0.108 & 1.01  & 0.500 & 0.108 & 1.01  & 0.500 & 0.108 & 1.01\\
$U_0$ only & 
0.471 & 0.116 & 0.96 & 0.484 & 0.116 & 0.95 & 0.477 & 0.112 & 1.02 \\
$U_0$ and $U_1$ & 
0.504 & 0.110 & 1.00 & 0.500 & 0.108 & 1.01 & 0.502 & 0.109 & 1.00  \\
$U_0$, $U_1$ and $U_2$ & 
0.500 & 0.108 & 1.01 & 0.500 & 0.108 & 1.01 & 0.499 & 0.108 & 1.01 
\end{tabular}
\caption{Calculated properties of the $K$-meson for various values of the
momentum partitioning parameter, $\eta_P$; ``$-$'' means that no bound state
solution exists in this case.  The mass $(m_K^{\rm exp}=0.496)$ and decay
constant ($f_K^{\rm exp}=0.113$) are in GeV, ${\cal R}_K$ is dimensionless.
\label{resK}}
\end{table}
\end{document}